\documentclass[aps,showpacs,pra,superscriptaddress,preprintnumbers,amsmath,amssymb,twocolumn,nolongbibliography]{revtex4-2}
\usepackage[utf8]{inputenc}
\usepackage[T1]{fontenc}
\usepackage{graphicx}
\usepackage{dcolumn}
\usepackage{appendix}
\usepackage{amsmath}
\usepackage{bm}
\usepackage[dvipsnames]{xcolor}
\usepackage{multirow}
\usepackage{notes2bib}
\usepackage{titletoc}
\usepackage{lipsum}
\usepackage{accents}

\usepackage{subfigure}
\usepackage{bbm}
\usepackage{enumerate}
\usepackage[
bookmarks=true,
colorlinks,
linkcolor=black,
urlcolor=black,
citecolor=black,
plainpages=false,
pdfpagelabels,
final,
breaklinks=true
]{hyperref}
\usepackage{titlesec}
\usepackage{array}
\usepackage{physics}

\newcommand{\relaxket}[1]{\lvert{#1}\rangle}
\newcommand{\relaxbra}[1]{\langle{#1}\rvert}

\begin{document}
\allowdisplaybreaks

\title{Microscopic analysis of above-threshold ionization driven by squeezed light}

\author{J.~Rivera-Dean}
\email{physics.jriveradean@proton.me}
\affiliation{ICFO -- Institut de Ciencies Fotoniques, The Barcelona Institute of Science and Technology, 08860 Castelldefels (Barcelona)}

\author{P.~Stammer}
\affiliation{ICFO -- Institut de Ciencies Fotoniques, The Barcelona Institute of Science and Technology, 08860 Castelldefels (Barcelona)}
\affiliation{Atominstitut, Technische Universität Wien, 1020 Vienna, Austria}

\author{C.~Figueira de Morisson Faria}
\affiliation{Department of Physics and Astronomy, University College London, Gower Street, London WC1E 6BT, UK}

\author{M. Lewenstein}
\affiliation{ICFO -- Institut de Ciencies Fotoniques, The Barcelona Institute of Science and Technology, 08860 Castelldefels (Barcelona)}
\affiliation{ICREA, Pg.~Llu\'{\i}s Companys 23, 08010 Barcelona, Spain}

\begin{abstract}
	Above-threshold ionization (ATI) is a strong-field-driven process where electrons absorb more photons than required for ionization.~While ATI dynamics and outputs are well-understood when driven by classical, perfectly coherent light, the recent development of non-classical light sources for strong-field phenomena has spurred interest in their effect on the involved electron dynamics.~In this work, we present a microscopic quantum optical theory describing ATI under the influence of strong squeezed light.~We observe that squeezed light significantly enhances the coupling between light and matter, making their mutual backaction more important than under classical driving.~This backaction profoundly impacts the electronic ionization times, as well as the non-classical properties of the joint electron-light state. This results in pronounced entanglement features, both immediately after ionization, and at later times. These entanglement features are reflected in the properties of the quantum optical state of the driving field revealing notable non-Gaussian features that depend on both, the amount of squeezing, and the number of ionization events occurring during the interaction.
\end{abstract}

\maketitle

\section{INTRODUCTION}
Strong-field physics investigates light-matter interactions in regimes where the intensity of the driving field becomes comparable to the atomic force binding electrons to their nuclei.~In these extreme conditions, and particularly when the frequency of the driving field is insufficient to ionize the atom via single-photon absorption (which is typical for fields in the infrared regime), a variety of nonperturbative phenomena can arise~\cite{keldysh_ionization_1965,amini_symphony_2019}, among which Above-Threshold Ionization (ATI) is a prominent example~\cite{milosevic_above-threshold_2006,lewenstein_principles_2008,agostini_chapter_2012}.~ATI refers to the ionization of an electron with absorption of more photons than required to surpass the atom's ionization potential. Under especially intense fields, this energy excess may correspond not to just a few photons~\cite{agostini_free-free_1979}, but to dozens or even hundreds~\cite{chin_observation_1983,hansch_resonant_1997}.~In such regimes, ionization is no longer adequately described as a multiphoton absorption process~\cite{voronov_ionization_1965,voronov_many_1966}, but rather as optical tunneling~\cite{chin_observation_1983}, where the laser field distorts the atomic potential forming a barrier through which the electron can tunnel and subsequently be accelerated by the field~\cite{keldysh_ionization_1965,lewenstein_rings_1995}.

\begin{figure}
	\centering
	\includegraphics[width=1\columnwidth]{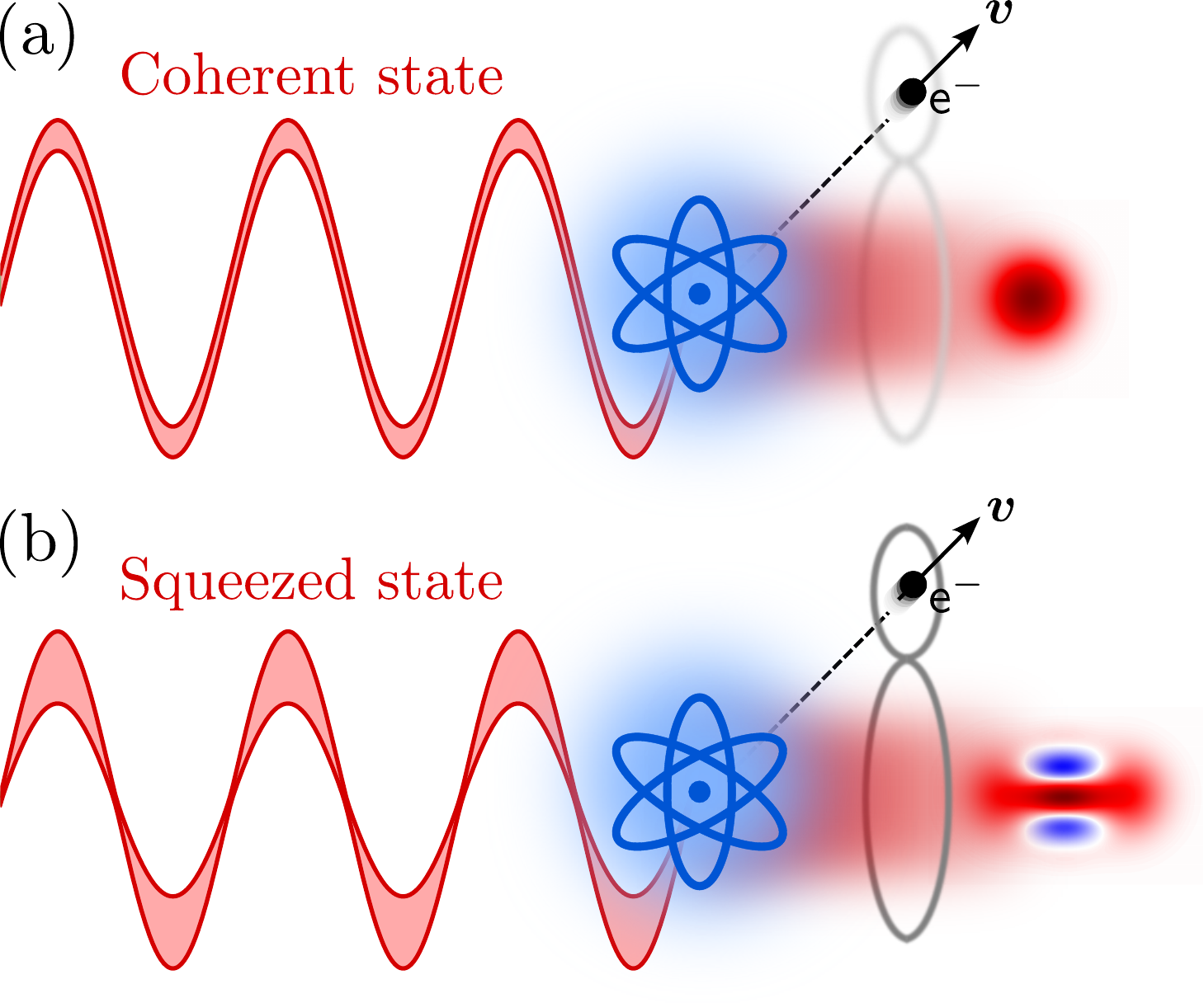}
	\caption{(a) When driven by coherent states of light, the backaction on the driving field from photoionized electrons generated in ATI processes is negligible, as are any non-classical features arising from the light-matter interaction. (b) By introducing squeezing into the driving field, these non-classical features become more pronounced, leading to non-negligible entanglement between electron and light. This entanglement can be exploited to engineer non-Gaussian states in the driving field.}
	\label{Fig:Scheme}
\end{figure}

Thus, from a semiclassical perspective---where matter is treated quantum mechanically and the electromagnetic field as a classical wave---the dynamics underlying ATI are well-understood across various levels of depth~\cite{lewenstein_rings_1995,lohr_above-threshold_1997,becker_above-threshold_2002,lai_influence_2015,maxwell_quantum_2015,maxwell_controlling_2016,becker_plateau_2018}.~However, the introduction of quantum optical frameworks has not only historically reinforced the theoretical foundations of ATI~\cite{guo_quantum_1988,guo_scattering_1989,aberg_scattering-theoretical_1991,guo_multiphoton_1992,fu_interrelation_2001,wang_frequency-domain_2007,wang_frequency-domain_2012}, but more recent developments over the last five years have enabled the prediction of phenomena beyond the reach of semiclassical models by explicitly treating the electromagnetic field as quantized, even when the light source is a conventionally classical laser~\cite{rivera-dean_light-matter_2022,stammer_quantum_2023,milosevic_quantum_2023,cruz-rodriguez_quantum_2024}.~In this direction, Ref.~\cite{milosevic_quantum_2023} developed a quantum theory of photoemission in ATI, predicting emission probabilities that are orders of magnitude higher than those associated with high-harmonic generation (HHG).~This enhancement stems from the single-step nature of direct ATI processes, as opposed to the more intricate three-step mechanism characteristic of HHG~\cite{krause_high-order_1992,corkum_plasma_1993,lewenstein_theory_1994}.~Additionally, Ref.~\cite{stammer_quantum_2023} found that the electron dynamics after ionization can impact the quantum state of the field, these effects resulting in mild yet non-negligible entanglement between the ionized photoelectrons and the driving field~\cite{rivera-dean_light-matter_2022}, as illustrated in Fig.~\ref{Fig:Scheme}~(a).

The use of quantum optical descriptions has also opened new avenues for understanding how non-classical states of light, such as squeezed states of light~\cite{walls_squeezed_1983,ScullyBookCh2}, can influence the electron dynamics in ATI. In an initial theoretical study, Ref.~\cite{balybin_photoionization_2019} demonstrated that employing squeezed light enhances the contribution of higher-order multiphoton ionization channels, and leads to the broadening of the peaks observed in photoelectron spectra.~More recently, analogous investigations in the optical tunneling regime have been conducted, utilizing a formalism originally developed for HHG~\cite{gorlach_high-harmonic_2023,even_tzur_photon-statistics_2023,stammer_absence_2024,rivera-dean_non-classicality_2025,gothelf_high-order_2025} based on the generalized positive $P$-representation~\cite{drummond_generalised_1980} to describe non-classical driving fields.~These studies confirmed the persistence of the previously observed effects: broadening of the ATI photoelectron peaks~\cite{fang_strong-field_2023,lyu_effect_2025}, extended ATI cutoffs~\cite{lyu_effect_2025,wang_high-order_2023}, and increased probability of different photoionization channels~\cite{liu_atomic_2025}, all attributed to the amplified field fluctuations characteristic of squeezed light. Importantly, recent experimental progress at this new intersection---driving metal needle tips with strong squeezed light---has shown how the photon statistics of the driving field are reflected in the measured photoelectron statistics within the multiphoton regime~\cite{heimerl_multiphoton_2024}, and has also demonstrated significant modifications in the photoelectron spectra in the tunneling regime~\cite{heimerl_driving_2025}.

These studies have provided valuable insights into how non-classical driving fields can modify standard ATI observables.~Yet, it remains open to understand how the use of non-classical drivers affects the joint quantum state of light and matter.~The aim of this work is to address this question by presenting a microscopic theory of light-matter interaction during ATI in the presence of squeezed light.~We show that, under relatively strong squeezing conditions---compatible with current state-of-art capabilities~\cite{manceau_indefinite-mean_2019,rasputnyi_high_2024,tzur_measuring_2025,lemieux_photon_2024}---the squeezing enhances the light-matter coupling, which in turn significantly affects the mutual backaction between light and matter.~This sets our theory apart from standard approximations compatible with classical, coherent driving fields where such backaction is treated only in one direction~\cite{lewenstein_generation_2021, rivera-dean_strong_2022, stammer_high_2022, rivera-dean_light-matter_2022,stammer_quantum_2023,rivera-dean_role_2024, stammer_theory_2025} or through Markov-like approximations~\cite{stammer_squeezing_2023,rivera-dean_squeezed_2024}.~It is worth noting that, in the context of HHG in strongly correlated materials driven by classical coherent fields, more comprehensive numerical studies have been conducted beyond these approximations~\cite{lange_electron-correlation-induced_2024,lange_excitonic_2025}, identifying the regimes in which such simplifications hold~\cite{lange_hierarchy_2025}.~Recently, similar analyses have been done for HHG in the context of sideband high-harmonic generation~\cite{boroumand_quantum_2025}.~As a consequence of the enhanced light-matter coupling induced by squeezing, we observe pronounced non-classical features~\cite{kenfack_negativity_2004,walschaers_non-gaussian_2021} in the quantum state of the driving field, as well as strong entanglement between light and matter, as illustrated in Fig.~\ref{Fig:Scheme}~(b).

\section{THEORETICAL ANALYSIS}
In this section, we present the theoretical framework adopted in this work, highlighting the key differences with respect to previous microscopic analyses of ATI in a quantum optical context~\cite{rivera-dean_light-matter_2022,stammer_quantum_2023}.

\subsection{Light-matter interaction Hamiltonian}
The primary objective of this work is to describe the interaction of a high-intensity displaced squeezed state with frequency $\omega_L$ and an atomic system that is initially in its ground state [Fig.~\ref{Fig:Scheme}~(b)].~Accordingly, we express the initial state of the joint light-matter system as
\begin{equation}
	\ket{\Psi(t_0)}
		= \ket{\text{g}} \otimes \ket{\xi,\alpha} \bigotimes_{q\neq 1} \ket{0_q}, 
\end{equation}
where $\ket{\text{g}}$ denotes the atomic ground state, and $\ket{\xi, \alpha} = \hat{D}_q(\alpha)\hat{S}_q(\xi)\ket{0}$ represents a displaced squeezed vacuum (DSV) state in the driving field mode $(q=1)$.~Here $\hat{D}_q(\alpha) = \text{exp}[\alpha \hat{a}_q^\dagger - \alpha^* \hat{a}_q]$ is the displacement operator, and $\hat{S}_q(\xi) = \text{exp}[\xi^* \hat{a}_q^{\dagger2}-\xi\hat{a}_q^2]$ the squeezing operator, with $\hat{a}_q$ ($\hat{a}^\dagger_q$) denoting the annihilation (creation) operator acting on the $q$th harmonic mode.~All other optical modes $(q > 1)$ are initially in the vacuum state $\relaxket{0_q}$. 

For DSV states, the mean photon number is given by $\langle \hat{a}_q^\dagger \hat{a}_q\rangle = \lvert \alpha\rvert^2 + \sinh[2](\abs{\xi})$. In this work, we focus on scenarios where, even in the absence of squeezing ($\xi = 0$), the coherent contribution ($\alpha$) alone is sufficient to induce strong-field dynamics in atomic systems~\cite{amini_symphony_2019}, corresponding to peak intensities $I_L \approx 10^{14}$ W/cm$^2$.~Within this framework, we investigate how the inclusion of squeezing affects both the ATI electron dynamics and the final state of the joint light-matter system.

Under the single-active electron and dipole approximation, the Hamiltonian describing the interaction between an atomic system and a quantized electromagnetic field can be expressed, within the length gauge~\cite{stammer_quantum_2023}, as 
\begin{equation}\label{Eq:Init:Hamiltonian}
	\hat{H} =
		\hat{H}_{\text{at}} + \mathsf{e} \hat{r} \hat{E} + \hat{H}_{\text{field}},
\end{equation}
where the light-matter interaction dynamics is governed by $i\hbar \pdv*{\ket{\Psi(t)}}{t} = \hat{H}\ket{\Psi(t)}$.~In Eq.~\eqref{Eq:Init:Hamiltonian}, $\hat{H}_{\text{at}} = \hat{p}^2/(2m_{\mathsf{e}})+V_{\text{at}}(\hat{r})$ denotes the atomic Hamiltonian, $\hat{E} = \sum_q \hat{E}_q= -i\sum_{q} g(\omega_q)[\hat{a}_q - \hat{a}^\dagger_q]$ is the electric field operator, and $\hat{H}_{\text{field}} = \sum_{q} \hbar \omega_q \hat{a}_q^\dagger \hat{a}_q$ represents the free-field Hamiltonian.~Throughout this work, we restrict ourselves to a discrete set of optical modes corresponding to harmonic orders of the fundamental frequency $\omega_L$ ($q=1\equiv L$), i.e., $\omega_q = q \omega_L$.~Besides simplifying the analytical treatment, this choice ensures finite values of the light-matter coupling $g(\omega_q)$, which depends on the quantization volume~\cite{ScullyBookCh1}. In standard strong-field physics parameter regimes, $g(\omega_q)$ is estimated to be on the order of $10^{-8}$ a.u.~\cite{rivera-dean_light-matter_2022}, leading to electric field amplitudes for the driving mode on the order of $\langle \alpha \vert \hat{E}_L \vert \alpha \rangle \propto 10^{7}$ V/m.~While this discrete-mode treatment is particularly suitable for ATI, which primarily involves interactions between the driving field mode and the atomic system, the use of a continuous light spectrum is shown to not significantly alter the results~\cite{stammer_quantum_2023}.

To describe the light-matter interaction dynamics, it is particularly convenient to work in the interaction picture with respect to $\hat{H}_{\text{field}}$,  which effectively transforms $\hat{a}_q \to \hat{a}_q e^{-i\omega_q t}$, rendering the electric field operator explicitly time-dependent.~Additionally, to leverage insights from strong-field semiclassical analyses~\cite{lewenstein_theory_1994}, it is useful to adopt a displaced frame with respect to the coherent state amplitude of the driving mode.~This transformation effectively modifies the Hamiltonian as
\begin{equation}
	\hat{H}(t) 
		= \hat{H}_{\text{at}}
			 + \mathsf{e} \hat{r}
			 		\big[
			 			E_{\text{cl}}(t) + \hat{E}(t)
			 		\big],
\end{equation} 
where $E_{\text{cl}}(t) = \langle\alpha\vert\hat{E}_L(t)\vert\alpha\rangle$. In this displaced frame, the initial state is given by $\lvert\bar{\Psi}(t_0)\rangle = \ket{\text{g}}\otimes \ket{\xi,0}\bigotimes_{q\neq1} \ket{0_q}$ and, at an arbitrary time $t$, it is related to the state in the original frame by $\relaxket{\Psi(t)} = e^{-i\hat{H}_{\text{field}}(t-t_0)/\hbar}\hat{D}_L(\alpha)\lvert\bar{\Psi}(t)\rangle$.

These unitary transformations are standard in quantum optical analyses of strong-field-driven interactions~\cite{gorlach_quantum-optical_2020,lewenstein_generation_2021,rivera-dean_strong_2022,rivera-dean_light-matter_2022,stammer_quantum_2023,rivera-dean_role_2024}, particularly in scenarios where the driving field is a coherent state.~Here, we take an additional step further by performing a transformation analogous to the displacement before but involving the squeezing operation $\hat{S}_L(\xi)$.~Specifically, we define $\lvert \bar{\Psi}(t)\rangle = \hat{S}_L(\xi) \lvert \Tilde{\Psi}(t)\rangle$, such that the initial state simplifies to $\lvert \Tilde{\Psi}(t_0)\rangle = \ket{\text{g}} \otimes \lvert \bar{0}\rangle$, where $ \lvert \bar{0}\rangle$ is the vacuum state in all modes.~Under this transformation, the electric field operator transforms to (see Appendix~\ref{Sec:App:Hamiltonian})
\begin{align}
	\hat{S}_L^\dagger(\xi)\hat{E}(t)\hat{S}_L(\xi)
		&= -i \big[
					f(\xi,t) \hat{a}_L
					-f^*(\xi,t)\hat{a}_L^\dagger
				\big]
			+ \sum_{q>1}\hat{E}_q(t) \nonumber
		\\&\equiv \hat{E}_{L}(\xi,t) + \hat{E}_{\text{uv}}(t),
\end{align}
where $f(\xi,t) = g(\omega_L)[\cosh(r)e^{-i\omega_L t} + \sinh(r)e^{i(\omega_L t - \theta)}]$, with $\xi = re^{i\theta}$ ($r > 0$).~Consequently, the effective Hamiltonian reads
\begin{equation}\label{Eq:effective:Hamiltonian}
	\hat{H}_{\text{eff}}(t)
		= \hat{H}_{\text{at}} 
			+ \mathsf{e} \hat{r}
				\big[
					E_{\text{cl}}(t) + \hat{E}_{L}(\xi,t) + \hat{E}_{\text{uv}}(t)
				\big].
\end{equation}

One of the main differences compared to the case where only classical coherent light drives the strong-field dynamics appears already at this stage: the light-matter coupling $g(\omega_L)$ becomes exponentially enhanced with the amount of squeezing $r$.~Thus, under the presence of substantial squeezing, the contribution of the quantum fluctuations $\mathsf{e}\hat{r}\hat{E}_{L}(\xi,t)$ can no longer be treated as a perturbation relative to the classical driving field $\mathsf{e}\hat{r}E_{\text{cl}}(t)$, as is typically the case with coherent state drivers.~Consequently, standard approximations commonly employed in quantum optical analyses of strong-field interactions---such as neglecting the backaction of the quantum optical state on the electron dynamics~\cite{lewenstein_generation_2021,rivera-dean_strong_2022,rivera-dean_light-matter_2022,stammer_quantum_2023,stammer_squeezing_2023,rivera-dean_role_2024,rivera-dean_squeezed_2024}---become inadequate in the regimes of interest here.

Interestingly, the stronger coupling can amplify non-classical behaviors of the post-interaction state~\cite{rivera-dean_light-matter_2022}.~This has been recently observed in Ref.~\cite{yi_generation_2024}, where enhanced light-matter coupling---achieved through the use of an optical cavity---led to non-classical features in the quantum optical state after HHG processes.~Moreover, the electron dynamics themselves can be substantially altered by these enhanced quantum fluctuations~\cite{even_tzur_photon-statistics_2023,rivera-dean_nonclassical_2024}.~It is important to note, however, that this enhancement only affects, directly, the driving field mode; all other harmonic orders remain perturbatively coupled, as they are not influenced by squeezing, though indirectly~\cite{gorlach_high-harmonic_2023,even_tzur_photon-statistics_2023,rivera-dean_non-classicality_2025,tzur_measuring_2025,lemieux_photon_2024,boroumand_quantum_2025}.~In this regard, Ref.~\cite{wang_high_2024} explored a complementary scenario, investigating HHG when selected harmonic modes were prepared in squeezed vacuum states.

\subsection{Light-matter interaction dynamics}
We now focus our attention on the dynamics governed by the Hamiltonian in Eq.~\eqref{Eq:effective:Hamiltonian}.~These dynamics are encapsulated by the time-evolution operator $\hat{U}(t)$, which satisfies
\begin{equation}
	i\hbar \pdv{\hat{U}(t)}{t}
		= \hat{H}_{\text{eff}}(t) \hat{U}(t).
\end{equation}
When applied to an initial state $\relaxket{\tilde{\Psi}(t_0)}$, this propagator yields the evolved state at any later time $t$, i.e., $\relaxket{\tilde{\Psi}(t)} = \hat{U}(t,t_0)\relaxket{\tilde{\Psi}(t_0)}$.~A general solution to this differential equation can be expressed via the following integral form~\cite{smirnova_anatomy_2007}
\begin{equation}\label{Eq:Gen:Prop}
	\hat{U}(t)
		= \hat{U}_0(t,t_0)
			- \dfrac{i}{\hbar}
				\int^t_{t_0}
					\dd t_1
						\hat{U}(t,t_1)
							\hat{V}(t_1)
								\hat{U}_0(t_1,t_0),
\end{equation}
where the Hamiltonian $\hat{H}_{\text{eff}}$ has been partitioned into two parts, $\hat{H}_0(t)$ and $\hat{V}(t)$, such that $i\hbar \partial\hat{U}_0(t)/\partial t = \hat{H}_0(t)\hat{U}_0(t)$.~The choice of how to partition the Hamiltonian---what to include in $\hat{H}_0(t)$ versus $\hat{V}(t)$---is not unique, it typically reflects the physical process under study or is guided by the aim to simplify the resulting equations.~Moreover, in different recursive iterations of Eq.~\eqref{Eq:Gen:Prop}, one may adopt different partitions to better isolate specific physical mechanisms of interest~\cite{lohr_above-threshold_1997,rivera_dean_quantum-optical_2019}.

Here, following Ref.~\cite{rivera_dean_quantum-optical_2019}, we consider a total of two recursive iterations of Eq.~\eqref{Eq:Gen:Prop}, developed in more detail in Appendix~\ref{Sec:App:TDSE}.~For the first iteration, we adopt the partition $\hat{H}^{(1)}_0(t) = \hat{H}_{\text{at}}$ and $\hat{V}^{(1)}(t) = \mathsf{e}\hat{r}[E_{\text{cl}}(t) + \hat{E}_L(\xi,t)+\hat{E}_{\text{uv}}(t)]$.~For the second recursive iteration, we instead use $\hat{H}^{(2)}_0(t) = \hat{H}_{\text{at}} + \mathsf{e}\hat{r}[E_{\text{cl}}(t) + \hat{E}_L(\xi,t)]$ and $\hat{V}^{(2)}(t) = \mathsf{e}\hat{r}\hat{E}_{\text{uv}}(t)$.~With this choice, we express Eq.~\eqref{Eq:Prop:Cont} as
\begin{align}
	\hat{U}(t)
		&= \hat{U}^{(1)}_0(t,t_0) \label{Eq:prop:zeroth}
			\\&\quad - \dfrac{i}{\hbar}
				\int^t_{t_0}\! \dd t_1
					\hat{U}^{(2)}_0(t,t_1)
						\hat{V}^{(1)}(t_1)
							\hat{U}_0^{(1)}(t_1,t_0)\label{Eq:prop:first}
		\\&\quad + \dfrac{1}{\hbar^2}
			\int^{t}_{t_0}\! \dd t_2 \!\int^{t_2}_{t_0}\!\dd t_1
				 \hat{U}(t,t_2)
				 	\hat{V}^{(2)}(t_2)
				 		\hat{U}_0^{(2)}(t_2,t_1)\nonumber
				 			\\&\hspace{3cm}\quad
				 			\times
				 			\hat{V}^{(1)}(t_1)
				 				\hat{U}_0^{(1)}(t_1,t_0)\label{Eq:prop:second},
\end{align}
with $i\hbar \partial{\hat{U}^{(i)}_0(t)}/\partial t = \hat{H}_0^{(i)}(t)\hat{U}^{(i)}_0(t)$.~In this formulation, the zeroth-order term [Eq.~\eqref{Eq:prop:zeroth}] corresponds to an evolution solely under the atomic Hamiltonian.~The first-order term [Eq.~\eqref{Eq:prop:first}] describes an intermediate transition at time $t_1$ mediated by the dipole interaction coupling with both the classical field contribution $E_{\text{cl}}(t)$ and the quantum optical fields $\hat{E}_L(\xi,t)$ and $\hat{E}_{\text{uv}}(t)$, followed by evolution driven by $H_0^{(2)}(t)$ until the final time $t$.~As we will see, this term allows us to capture the dynamics underlying ATI.~Finally, and on top of the dynamics described by Eq.~\eqref{Eq:prop:first}, the second-order term [Eq.~\eqref{Eq:prop:second}] introduces an additional transition at time $t_2$, mediated exclusively by the dipolar coupling to all optical modes distinct from the driving field.~This contribution therefore accounts for the HHG dynamics.

\subsubsection{The Strong-Field Approximation}
A connection between the physical processes described by Eqs.~\eqref{Eq:prop:zeroth}-\eqref{Eq:prop:second} and the strong-field dynamics becomes more transparent upon introducing the strong-field approximation (SFA) in its standard formulation~\cite{lewenstein_theory_1994,lewenstein_rings_1995,amini_symphony_2019} (see Appendix~\ref{Sec:App:SFA:ATI} for a detailed analysis).~Within the SFA framework, the following assumptions are made:
\begin{enumerate}
	\item The strong laser field couples exclusively to the (non-degenerate) ground state $\ket{\text{g}}$, and not to the bound states.~As a result, the dynamics are confined between the ground state and the continuum states $\{\ket{k}\}$.
	\item Once ionized, the electron is treated as a free particle evolving solely under the influence of the external electric field.~The interaction with the nuclear potential can be incorporated as a perturbative correction to the electronic motion.
\end{enumerate}

The first of these assumptions holds whenever there are no intermediate resonances in the atomic system and when tunneling ionization is the dominant ionization channel.~The latter condition is satisfied in regimes where the Keldysh parameter $\gamma = \sqrt{2I_pm_{\mathsf{e}}\omega^2/(\mathsf{e}^2E_0^2)}\lesssim 1$~\cite{lewenstein_theory_1994,amini_symphony_2019}, with $E_0$ denoting the electric field amplitude.~Under these conditions, we can introduce an SFA-version of the identity over the atomic Hilbert space as
\begin{equation}\label{Eq:SFA:identity}
	\mathbbm{1}	
		= \dyad{\text{g}}
			+	\int \dd k \dyad{k},
\end{equation}
which, when inserted after every $\hat{V}^{(i)}(t_i)$ term in Eqs.~\eqref{Eq:prop:first} and \eqref{Eq:prop:second}, enables us to identify the first-order term with ATI processes, and the second-order term---though not exclusively---to the HHG process (see Appendix~\ref{Sec:App:SFA:ATI}). Therefore, by applying the time-evolution operator $\hat{U}(t,t_0)$ to the initial state and using Eq.~\eqref{Eq:SFA:identity}, we obtain
\begin{equation}\label{Eq:Total:state}
	\relaxket{\tilde{\Psi}(t)}
		\simeq \relaxket{\tilde{\Psi}_0(t)}
			+ \relaxket{\tilde{\Psi}_{\text{ATI}}(t)}
			+ \relaxket{\tilde{\Psi}_{\text{HHG}}(t)},
\end{equation}
where $\relaxket{\tilde{\Psi}_0(t)}$, corresponding to Eq.~\eqref{Eq:prop:zeroth}, describes the contribution where the electron remains in the ground state up to time $t$; $\relaxket{\tilde{\Psi}_{\text{ATI}}(t)}$, corresponding to Eq.~\eqref{Eq:prop:first}, accounts for ATI processes; and $\relaxket{\tilde{\Psi}_{\text{HHG}}(t)}$, corresponding to Eq.~\eqref{Eq:prop:second}, for HHG processes.

In this work, we focus on describing ATI events under the influence of strong squeezed light, which we later on isolate by performing suitable projective operations.~Accordingly, we restrict our attention to the ATI component of the wavefunction, $\relaxket{\tilde{\Psi}_{\text{ATI}}(t)}$, which can be expressed as (see Appendices~\ref{Sec:App:TDSE} and \ref{Sec:App:SFA:ATI})
\begin{equation}\label{Eq:ATI:state}
	\begin{aligned}
	\relaxket{\Tilde{\Psi}_{\text{ATI}}(t)}
		= - \dfrac{i\mathsf{e}}{\hbar}
			&\int^t_{t_0} \dd t_1
				\int \dd k\
					\hat{U}_L(t,t_1)
						\langle k\vert \hat{r}\vert \text{g}\rangle
						\\&\quad\times
						\big[
							E_{\text{cl}}(t_1)
							+ \hat{E}_L(\xi,t_1)
							+ \hat{E}_{\text{uv}}(t_1)
						\big]
						\\&\quad\times
						e^{iI_p(t_1-t_0)/\hbar}
						\ket{k}\otimes \ket{\bar{0}},
	\end{aligned}
\end{equation}
where $I_p$ is the ionization potential, defined via $\hat{H}_{\text{at}}\ket{\text{g}} = -I_p \ket{\text{g}}$, and $\hat{U}_L(t)$ is a time-evolution operator satisfying
\begin{equation}\label{Eq:Prop:Cont}
	i\hbar \pdv{\hat{U}_L(t)}{t}
		= \bigg[
				\dfrac{\hat{p}^2}{2m}
				+ V_{\text{at}}(\hat{r})
				+ \mathsf{e} \hat{r}
					\big(
						E_{\text{cl}}(t)
						+ \hat{E}_L(\xi,t)
					\big)
			\bigg] \hat{U}_L(t).
\end{equation}

Equation~\eqref{Eq:ATI:state} captures the sequence of steps involved in an ATI event. From the initial time $t_0$ until $t_1$, the electron remains in the ground state, accumulating a phase proportional to the ionization potential $I_p$.~At time $t_1$, it transitions into a continuum state $\ket{k}$ via a dipolar interaction with the field.~From $t_1$ to the final time $t$, the electron propagates according to Eq.~\eqref{Eq:Prop:Cont}, which governs its evolution under the combined influence of the classical driving field, its quantum fluctuations, and the atomic potential.~Within the SFA framework, this expression accounts for both direct and high-order ATI processes: in direct ATI (dATI), the electron gets ionized and escapes without returning to the atomic core, while in high-order ATI (HATI), the electron undergoes elastic recollisions with the parent ion, reaching higher continuum energies that are inaccessible through direct ATI alone~\cite{lewenstein_rings_1995,amini_symphony_2019}.

\subsubsection{The Direct ATI component}
For the remainder of our analysis, we focus on dATI events, where the electron propagates far from the nucleus while populating high-energetic continuum states. In this regime, and accordingly to the second of the SFA assumptions introduced earlier, the Coulomb potential can be treated perturbatively~\cite{lewenstein_theory_1994,lewenstein_rings_1995}, yielding
\begin{equation}\label{Eq:perb:theory:ATI}
	\relaxket{\Tilde{\Psi}_{\text{ATI}}(t)}
		\approx 
			\relaxket{\Tilde{\Psi}_{\text{d}}(t)}
			+ \relaxket{\Tilde{\Psi}_{\text{h}}(t)}
\end{equation}
where the zeroth-order term accounts for dATI events, and higher-order terms capture HATI contributions~\cite{lewenstein_rings_1995,amini_symphony_2019}, denoted with ``d'' and ``h'' subscripts, respectively.~This approximation is particularly appropriate for the processes considered here, in which typical photoelectron energies lie below $2U_p$, with $U_p = E_0^2/(4\omega_L^2)$ the pondemorotive energy.~More specifically, it performs well when working with strong-laser fields in the near-infrared regime ($\lambda _L \approx 800$ nm), though with some caveats~\cite{maxwell_coulomb-corrected_2017}.~However, it becomes less accurate in the mid-infrared regime ($\lambda_L \approx 2000$ nm), where the electron is more slowly driven away from the ion, resulting in pronounced low energy structured in the photoelectron spectrum~\cite{quan_classical_2009,blaga_strong-field_2009,yan_low-energy_2010}, which are not captured by the standard SFA. 

Thus, the evolution governed by the zeroth-order term in perturbation theory around the atomic potential reads, for a generic state $\ket{\psi(t)}$,
\begin{equation}\label{Eq:zeroth:perb:dyn}
	i\hbar \pdv{\ket{\psi(t)}}{t}
		= \bigg[
				\dfrac{\hat{p}^2}{2m}
				+ \mathsf{e} \hat{r}
					\big(
						E_{\text{cl}}(t)
						+ \hat{E}_L(\xi,t)
					\big)
		\bigg] \ket{\psi(t)},
\end{equation}
where we define $\ket{\psi(t)} = \hat{U}_L(t)\ket{\psi(t_0)}$, with $\ket{\psi(t_0)}$ an arbitrary initial state. A general solution to Eq.~\eqref{Eq:zeroth:perb:dyn} can be expressed as (see Appendix~\ref{Sec:App:SFA:ATI})
\begin{equation}\label{Eq:Psi:ev:continuum}
	\begin{aligned}
	\ket{\psi(t)}
		&= \hat{U}_{\text{vg}}(t)
				\hat{U}_V(t,t_0 )
					\hat{U}^\dagger_{\text{vg}}(t_0)
						\ket{\psi(t_0)}
		\\&\equiv \hat{U}_{\text{vg}}(t)
					\relaxket{\bar{\psi}(t)},
	\end{aligned}
\end{equation}
where $\hat{U}_{\text{vg}}(t) \equiv e^{i \mathsf{e}\hat{r}[A_{\text{cl}}(t) + \hat{A}_L(\xi,t)]/\hbar}$ denotes a gauge transformation---which in the limit $g(\omega_L)\to0$ corresponds to the standard length to velocity gauge transformation---and with $\relaxket{\bar{\psi}(t)}) = U_V(t,t_0)\relaxket{\bar{\psi}(t)} \relaxket{\psi(t_0)}$ satisfying
\begin{equation}
	i\hbar \pdv{\relaxket{\bar{\psi}(t)}}{t}
		= \dfrac{1}{2m}
				\Big[
					\hat{p} 
					+ \mathsf{e} A_{\text{cl}}(t)
					+ \mathsf{e} \hat{A}_L(\xi,t)
				\Big]^2
				\relaxket{\bar{\psi}(t)}.
\end{equation}
Here, $A_{\text{cl}}(t)$ and $\hat{A}_L(\xi,t)$ denote classical and quantum vector potentials, respectively, related to the corresponding electric fields via $E_{\text{cl}}(t) = - \partial{A_{\text{cl}}(t)}/{\partial t}$ and $\hat{E}_L(\xi,t) = - \partial{\hat{A}_L(\xi,t)}/{\partial t}$.

It is worth highlighting that the total time-evolution operator in Eq.~\eqref{Eq:Psi:ev:continuum} cannot, in general, be decomposed as $\hat{U}_{\mathsf{e}}(t,t_0)\otimes \hat{U}_{\text{field}}(t,t_0)$. This reflects the fact that, during the electron's excursion in the continuum, entanglement between light and matter naturally emerges.~The degree of this entanglement critically depends on the amount of squeezing; when $\cosh(r)g(\omega_L) \to 0$, Eq.~\eqref{Eq:Psi:ev:continuum} simplifies to $\hat{U}_{\mathsf{e}}(t,t_0)\otimes \mathbbm{1}$, and the evolution becomes separable: in this limit, our expressions coincide with those from semiclassical analyses~\cite{lewenstein_rings_1995,amini_symphony_2019}.~Here, we focus on regimes where $\abs{\cosh(r)g(\omega_L)} \leq \abs{\alpha g(\omega_L)}$, with the right-hand side representing the electric field amplitude.~That is, we consider scenarios in which the contribution from squeezing is, at most, comparable to that of the coherent component of the driving field, but never dominant.~This parameter regime is consistent with current experimental implementations~\cite{manceau_indefinite-mean_2019,rasputnyi_high_2024,tzur_measuring_2025,lemieux_photon_2024}, where the intensity associated with the squeezed component is typically two orders of magnitude smaller than that of the coherent component required to drive strong-field processes in atomic systems.

Recalling that, in the context of dATI events, the propagator can be approximated as $\hat{U}_L(t,t_1) \approx  \hat{U}_{\text{vg}}(t)\hat{U}_V(t,t_1 )\hat{U}^\dagger_{\text{vg}}(t_1)$, we insert this expression into Eq.~\eqref{Eq:ATI:state} to obtain an explicit form of the dATI component of the state
\begin{equation}\label{Eq:dATI:state}
	\begin{aligned}
		\relaxket{\Tilde{\Psi}_{\text{d}}(t)}
			&= - \dfrac{i\mathsf{e}}{\hbar}
		\int^t_{t_0} \dd t_1
			\int \dd k\
		\hat{U}_{\text{vg}}(t)
			\hat{U}_V(t,t_1)
				\hat{U}^\dagger_{\text{vg}}(t_1)
		\\&\quad\times 	\langle k\vert \hat{r}\vert \text{g}\rangle
			\big[
				E_{\text{cl}}(t_1)
				+ \hat{E}_L(\xi,t_1)
				+ \hat{E}_{\text{uv}}(t_1)
			\big]
		\\&\quad\times
			e^{iI_p(t_1-t_0)/\hbar}
				\ket{k}\otimes \ket{\bar{0}}.
	\end{aligned}
\end{equation}
In the following, we analyze the role of each term in the decomposition of $\hat{U}_L(t,t_1)$, and how they influence the light–matter interaction dynamics.

\subsubsection{Light-matter entanglement after ionization}\label{Sec:ion:ent}
The action of $\hat{U}_{\text{vg}}(t)$ is best understood by moving to the position representation for the electronic degrees of freedom.~In this framework, we find that 
\begin{equation}
	\hat{U}_{\text{vg}}(t) \ket{x}
		= e^{-i\mathsf{e} x A_{\text{cl}}(t)/\hbar}
				\hat{D}_L
					\big(
						\tfrac{\mathsf{e}}{\hbar}x F(\xi,t)
				\big)
					\ket{x},
\end{equation}
that is, $\hat{U}_{\text{vg}}(t)$ induces a displacement on the driving field mode that depends on the electron's position and the amount of squeezing through
\begin{equation}
	F(\xi,t)
		= i g(\omega_L)
			\bigg[
				\dfrac{\cosh(r)}{\omega}e^{i\omega t}
				- \dfrac{\sinh(r)}{\omega}e^{-i(\omega t-\theta)}
			\bigg].
\end{equation}

Therefore, by introducing the identity in the position represention for the electronic degrees of freedom, we can rewrite Eq.~\eqref{Eq:dATI:state} as
\begin{equation}\label{Eq:Psi:dATI:with:Psi:ion}
	\begin{aligned}
	\relaxket{\Tilde{\Psi}_{\text{d}}(t)}
		&= - \dfrac{i\mathsf{e}}{\hbar}
			\int^t_{t_0} \!\dd t_1
				\hat{U}_{\text{vg}}(t)
					\hat{U}_V(t,t_1)
					e^{\frac{iI_p}{\hbar}(t_1-t_0)}
					\relaxket{\tilde{\Psi}_{\text{ion}}(t_1)},
	\end{aligned}
\end{equation}
where $\relaxket{\Psi_{\text{ion}}(t_1)}$ represents the joint light-matter system state immediately after ionization, given by
\begin{equation}\label{Eq:Psi:ion}
	\begin{aligned}
	\relaxket{\Psi_{\text{ion}}(t_1)}
		&= \int \dd x 
				\hat{D}_L
					\big(
						\tfrac{\mathsf{e}}{\hbar}x F(\xi,t)
					\big)h(x)
				\\&\hspace{0.2cm}\times
				\big[
					E_{\text{cl}}(t_1)
					+ \hat{E}_L(\xi,t_1)
					+ \hat{E}_{\text{uv}}(t_1)
				\big]
				\ket{x}\otimes \ket{\bar{0}},
	\end{aligned}
\end{equation}
where $h(x) \equiv \int \dd k \braket{x}{k}\!\mel{k}{\hat{r}}{\text{g}}$.~In this expression, $h(x)$ provides the probability amplitude for the electron to tunnel through the potential barrier and emerge at position $x$. The displacement operator $\hat{D}_L(\mathsf{e}xF(\xi,t_1)/\hbar)$ encodes the effect of this tunneling event on the field degrees of freedom, resulting in a position-dependent displacement of the driving field mode.~Importantly, the magnitude of this displacement increases with the amount of squeezing, as the enhanced field fluctuations provide a broader range of quantum paths for the electron to tunnel through.

\begin{figure}
	\centering
	\includegraphics[width = 1\columnwidth]{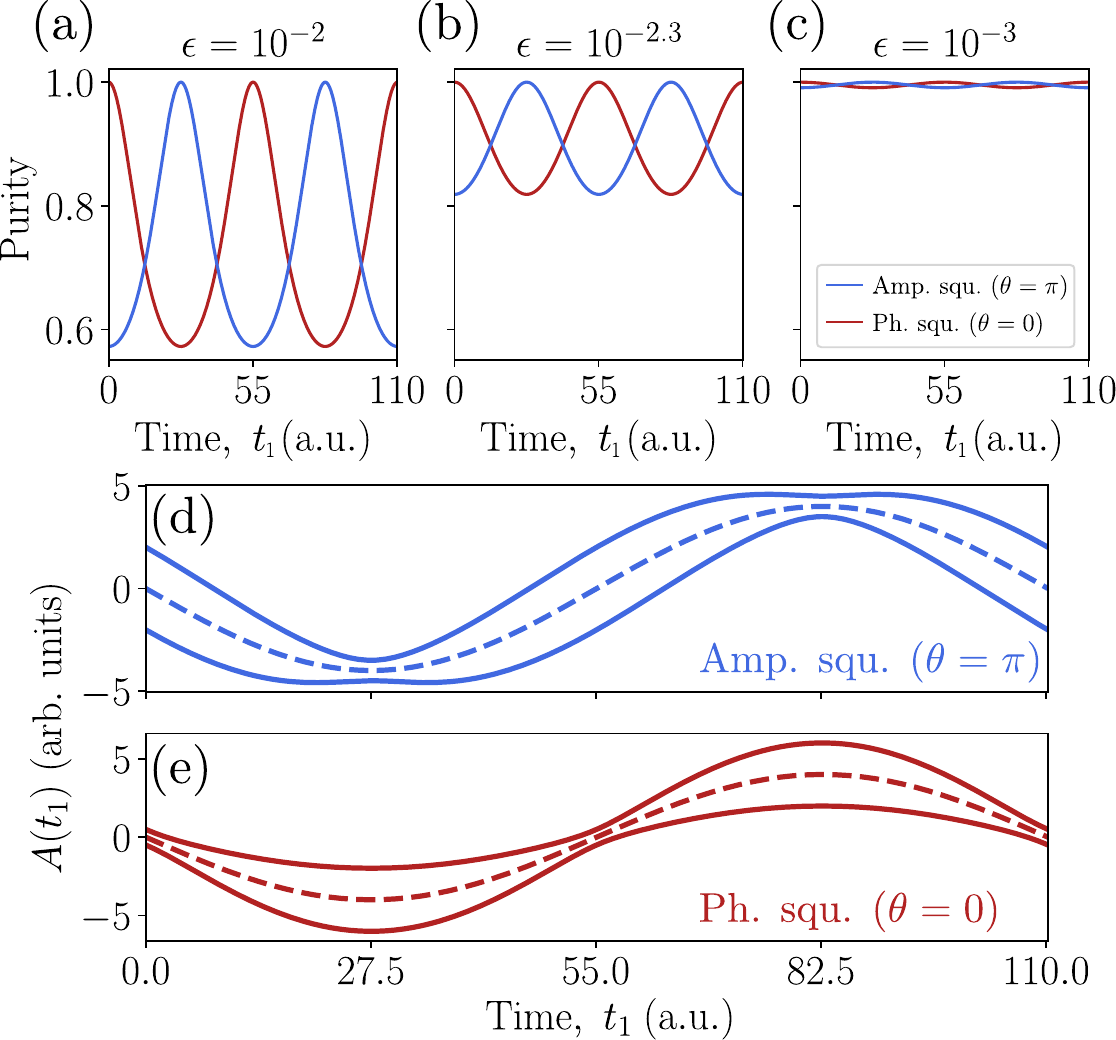}
	\caption{Panels (a)-(c) show the purity $\gamma = \tr[\hat{\rho}^2_{\text{field}}(t)]$, with $\hat{\rho}_{\text{field}}(t) = \tr_{\text{elec}}[\relaxket{\Tilde{\Psi}_{\text{ion}}(t_1)}\!\relaxbra{\Tilde{\Psi}_{\text{ion}}(t_1)}]$, computed at different times $t_1$ and for varying squeezing strengths, with $\epsilon = e^{r}g(\omega_L)$.~Results for amplitude squeezing ($\theta=\pi$) and phase squeezing ($\theta=0$) are shown in blue and red, respectively.~Panels (d) and (e) display, for reference, the expectation value (dashed curve) and fluctuations (solid curves) of the vector potential operator, evaluated with respect to amplitude-squeezed and phase-squeezed states, respectively.}
	\label{Fig:purity:ion}
\end{figure}

To gain further insight into the extent of light-matter entanglement in Eq.~\eqref{Eq:Psi:ion}, we consider the scenario in which the classical field component $E_{\text{cl}}(t_1)$ dominates. In this regime, we evaluate the degree of entanglement by analyzing
\begin{equation}
	\relaxket{\Psi_{\text{ion}}(t_1)}
	\approx \int \dd x 
		\hat{D}_L
			\big(
				\tfrac{\mathsf{e}}{\hbar}x F(\xi,t_1)
			\big)h(x)
	\ket{x}\otimes \ket{\bar{0}},
\end{equation}
and quantifying the state's purity after tracing out either the electronic or photonic degrees of freedom: since the full light-matter state remains pure due to unitary evolution, obtaining a mixed reduced state upon tracing out one subsystem indicates a loss of coherence arising from pre-existing quantum entanglement between the two.~The results are presented in Fig.~\ref{Fig:purity:ion}, where the purity is computed for both amplitude and phase squeezing, across varying squeezing strengths defined by $\epsilon \equiv e^r g(\omega_L)$ ($\cosh(r)\approx \sinh(r) \approx e^r$ for $r \gg 1$).~In both cases, the purity exhibits oscillations between two well-defined bounds, which converge as $\epsilon$ decreases.~This indicates that stronger squeezing leads to higher degrees of entanglement, with the precise value depending on the ionization time.~In the absence of squeezing, the amount of entanglement is minimal at all times.

Interestingly, we find that amplitude and phase squeezing yield purity oscillations that are out of phase by half-cycle, and these oscillations occur with twice the frequency of the driving vector potential (see panels (d) and (e)).~This behavior arises because the purity reaches its maximum at times when the vector potential fluctuations---responsible for determining the electron's kinetic energy upon tunneling---are themselves maximal. Larger field fluctuations result in increased spatial delocalization of the electron immediately after ionization, broadening the distribution of possible optical displacements in Eq.~\eqref{Eq:Psi:ion}, thereby enhancing entanglement. 

\subsubsection{Propagation in the continuum}
While $\hat{U}_{\text{vg}}(t_1)$ encodes light-matter correlations at the ionization time, $\hat{U}_{V}(t,t_1)$ accounts for correlations that develop during the electron's propagation from the ionization time $t_1$ to the final detection time $t$.~To analyze its contribution in Eq.~\eqref{Eq:Psi:dATI:with:Psi:ion}, it is particularly convenient to insert the identity in the momentum representation, allowing the state to be rewritten as
\begin{equation}
	\begin{aligned}
		\relaxket{\Tilde{\Psi}_{\text{d}}(t)}
			&= - \dfrac{i\mathsf{e}}{\hbar}
				\int^t_{t_0} \dd t_1 \int \dd v
					\ \hat{U}_{\text{vg}}(t)
						\hat{U}_{V}(t,t_1)\ket{v}
							\\&\hspace{2.5cm}\times
								e^{i\frac{I_p}{\hbar}(t_1-t_0)}
									\langle v \relaxket{\tilde{\Psi}_{\text{ion}}(t_1)}.
	\end{aligned}
\end{equation}
For the regime of squeezing parameters used in this work, we find in Appendix~\ref{Sec:App:SFA:ATI} that 
\begin{equation}
	\begin{aligned}
	\hat{U}_V(t,t_1)\ket{v}
		&\approx \exp[-\dfrac{i}{2m_{\mathsf{e}}\hbar}
								\int^t_{t_1}
									\dd \tau
										\big(
											p + \mathsf{e}A_{\text{cl}}(t_1)
										\big)^2]
						\\&\quad \times
						\hat{D}
							\big(
								\delta(v,t,t_1)
							\big)\ket{v},
	\end{aligned}
\end{equation}
where this expression reveals that the electron not only accumulates a propagation-dependent phase, but also induces an additional displacement on the quantum optical degrees of freedom, given by
\begin{equation}\label{Eq:disp:squ}
	\delta(v,t,t_0)
		= \dfrac{\mathsf{e}}{2m_{\mathsf{e}}\hbar}
			\int_{t_0}^t \dd \tau
				\big[
					v + A_{\text{cl}}(\tau)
				\big]F(\xi,\tau).
\end{equation}

This displacement has its analogue in Refs.~\cite{rivera-dean_light-matter_2022,rivera-dean_role_2024}, where the backation of the electron dynamics on the quantum optical degrees of freedom was analyzed for ATI and HHG processes driven by coherent state drivers, respectively.~In those cases, however, the resulting displacement was found to be proportional to
\begin{equation}\label{Eq:disp:old}
	\delta_{\scriptsize\cite{rivera-dean_light-matter_2022,rivera-dean_role_2024}}
		\propto 
			\int^{t}_{t_1} \dd t_2 
				\int^{t_2}_{t_1}\dd \tau
					\big[
						v_1 + \mathsf{e}A_{\text{cl}}(\tau)
					\big]
					e^{i\omega \tau},
\end{equation}
which does not reduce to Eq.~\eqref{Eq:disp:squ} when $r=0$.~The discrepancy can be traced back to the approximations made in Refs.~\cite{rivera-dean_light-matter_2022,rivera-dean_role_2024}.~A key assumption in that model was that the electron trajectories remained unperturbed by the quantum fluctuations---i.e., the field backaction on the electron was neglected---while retaining the backaction of the electron on the field.~Within this regime, the electron behaves as a classical charge current oscillating under the influence of the driving field, inducing a dipole moment proportional to $\int^t_{t_1} \dd \tau [v_1 + \mathsf{e}A_{\text{cl}}(\tau)]$, which in turn modulates the amplitude of the quantum optical modes in a mean-field-like manner. That is, the field responds to the electron motion, but not vice versa, with the effects being obtained through a Fourier transform of the effective charge current~\cite{ScullyBookCh2}.

\begin{figure}
	\centering
	\includegraphics[width=1\columnwidth]{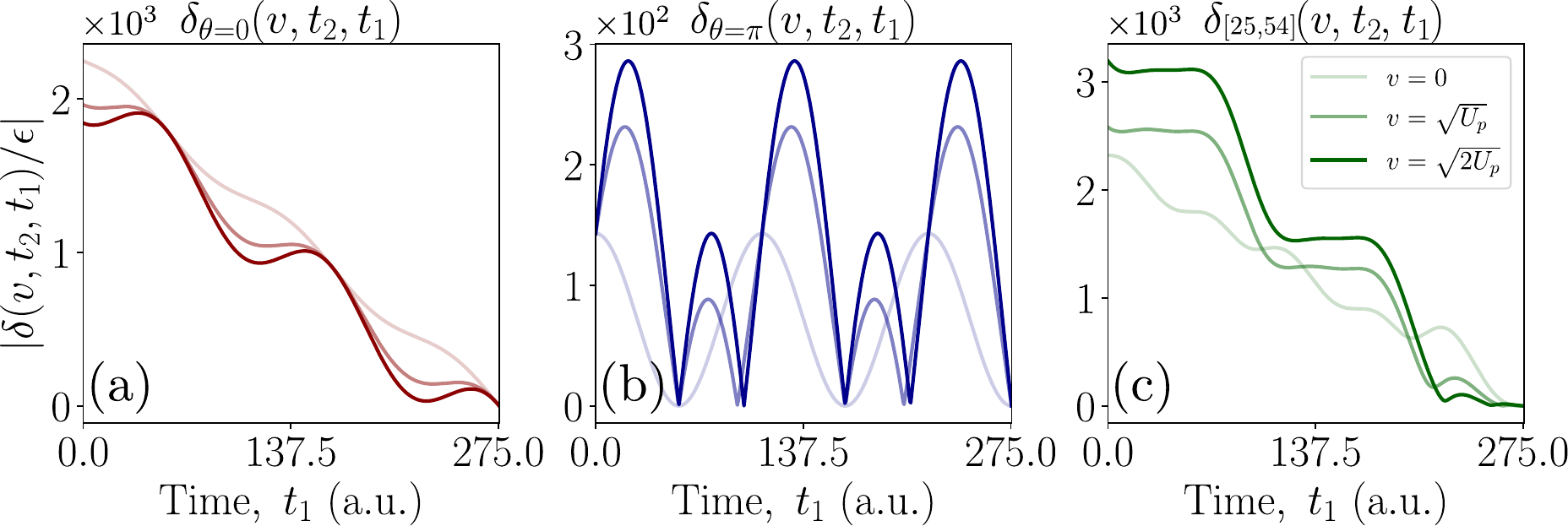}
	\caption{Comparison of the displacements given by Eq.~\eqref{Eq:disp:squ} (panel (a) for phase squeezing and panel (b) for amplitude squeezing) and Eq.~~\eqref{Eq:disp:old} (panel (c)) as a function of the ionization time $t_1$.~In all cases, we set $E_0 = 0.053$ a.u.~and $\omega_L = 0.057$ a.u., and fix $t_2 =5\pi/\omega_L$.~The displacement in each plot is normalized by the corresponding value $\epsilon$.}
	\label{Fig:deltas:comp}
\end{figure}

In contrast, the present work goes beyond this approximation by including strongly squeezed states of light, which exponentially enhance the light-matter coupling with the squeezing parameter.~As a result, the interaction during the electron's excursion in the continuum significantly affects both the field amplitude and the electron dynamics.~The increased coupling strength leads to a bidirectional feedback: the field's quantum fluctuations perturb the electron trajectories, while the electron's motion non-perturbatively displaces the field.~This mutual influence invalidates a mean-field description and highlights the inherently entangled nature of the light-matter interaction in the presence of squeezed drivers.

In Fig.~\ref{Fig:deltas:comp}, we compare the displacement given by Eq.~\eqref{Eq:disp:squ}, for phase and amplitude squeezing---shown in panels (a) and (b), respectively---with that of Eq.~\eqref{Eq:disp:old}, shown in panel (c).~All displacements are plotted in units of $\epsilon$ as a function of the ionization time $t_1$, with the final time fixed at $t_2 = 5\pi/\omega_L$.~For both phase and amplitude squeezing, we consider large values of the squeezing parameter $r$, such that $\cosh(r) \approx \sinh(r) \approx e^{r}$. As shown, the displacement behavior varies significantly across the three cases.~First, we observe that the displacement increases with earlier ionization times in the phase-squeezed and mean-field cases, but not for amplitude squeezing.~Additionally, for both the amplitude-squeezed and mean-field cases, the displacement tends to increase with the final electron momentum, whereas in the phase-squeezed case it remains relatively insensitive.~These differences highlight the limitations of the mean-field approximation underlying Eq.~\eqref{Eq:disp:old}, which neglects the mutual backaction between the electron and the field.

With all this in place, and by having in mind that the analysis of $\hat{U}^\dagger_{\text{vg}}(t)$ is alike that of $\hat{U}_{\text{vg}}(t)$ in Sec.~\ref{Sec:ion:ent}, the total quantum state after dATI events can be written as
\begin{widetext}
\begin{equation}\label{Eq:Psi:dATI:final}
	\begin{aligned}
		\relaxket{\Tilde{\Psi}_{\text{d}}(t)}
			&= - \dfrac{i\mathsf{e}}{\sqrt{2\pi\hbar^3}}
					\int^t_{t_0} \dd t_1
						\int \dd x_2
							\int \dd v
								\int \dd x_1
									\hat{D}_L
										\big(
											\alpha(v,t,t_1,x_2,x_1)
										\big)
									\big[
										E_{\text{cl}}(t_1)
										+ \hat{E}_L(\xi,t_1)
										+ \hat{E}_{\text{xuv}}(t)
									\big]
									\\& \hspace{4.7cm}\times
									h(x_1)
									e^{-i[
											S_{\text{sc}}(v,t,t_1)
											- x_2(v + \mathsf{e}A_{\text{cl}}(t))
											 + x_1(v + \mathsf{e}A_{\text{cl}}(t_1))]/\hbar}
									\ket{x_2}\otimes \ket{0},
	\end{aligned}
\end{equation}
\end{widetext}
where $S_{\text{sc}}(v,t,t_1) = \tfrac{1}{2m_{\mathsf{e}}} \int^{t}_{t_1}\dd \tau [p+\mathsf{e}A_{\text{cl}}(\tau)]^2 -\hbar I_p(t_1-t_0)$ denotes the semiclassical action and $\alpha(v,t,t_1,x_2,x_1)$ represents the total quantum optical displacement, given by
\begin{equation}\label{Eq:def:alpha}
	\alpha(v,t,t_1,x_2,x_1)
		= \delta(v,t,t_1)
			-	\tfrac{\mathsf{e}}{\hbar}x_2 F(\xi,t)
			+
				\tfrac{\mathsf{e}}{\hbar}x_1 F(\xi,t_1).
\end{equation}
We note that, due to the contribution of $\hat{U}^{\dagger}_{\text{vg}}(t)$, we get an additional displacement of $-\mathsf{e}x_2F(\xi,t)/\hbar$ on the driving field mode.~In what follows, we quantify the non-classical features of this state and analyze how the mutual backaction between the field and electronic degrees of freedom perturbs the electronic trajectories.

\section{RESULTS}
For the analysis presented in this section, we set $E_{\text{cl}}(t) = E_0 \cos(\omega_L t)$, with $\omega_L = 0.057$ a.u.~and $E_0 = 0.053$ a.u., corresponding to a wavelength $\lambda_L = 800$ nm and a peak intensity of $I_L=10^{14}$ W/cm$^2$.~As the atomic system, we consider a hydrogen atom with ionization potential $I_p = 0.5$ a.u., modeled using a Gaussian potential~\cite{lewenstein_theory_1994,nayak_saddle_2019}.~Furthermore, in what follows, we set the measurement time $t = 2\pi n_{\text{cyc}}/\omega$, where $n_{\text{cyc}}\in \mathbbm{W}$ denotes the number of field cycles.~This choice ensures that $A_{\text{cl}}(t) = 0$, so that the semiclassical kinetic and canonical momenta coincide instantaneously.

\subsection{Modification of the ionization times}
In analyses of HHG driven by squeezed light, it has been shown that the use of states with non-Poissonian photon statistics can induce an effective force that modifies the semiclassical electron trajectories~\cite{even_tzur_photon-statistics_2023}, even enabling HHG in scenarios that are otherwise forbidden when using coherent state drivers~\cite{rivera-dean_non-classicality_2025}. In those studies, the effective force was solely attributed to the photon statistics of the driver, as the mutual backaction between the light and matter resulting from their interaction was neglected~\cite{lewenstein_generation_2021,rivera-dean_strong_2022,stammer_quantum_2023}. In contrast, here we investigate how the ionization events leading to ATI are perturbed by this mutual interaction.

For that purpose, we follow an approach similar to that presented in Ref.~\cite{rivera-dean_role_2024}, and express the quantum optical state $\relaxket{\Tilde{\Phi}_{\text{d}}(v_{\text{f}},t)} \equiv \langle v_{\text{f}} \vert \Tilde{\Psi}_{\text{d}}(t)\rangle$, obtained by projecting the total state onto a final electronic momentum state $\ket{v_{\text{f}}}$, in the Fock basis as
\begin{equation}\label{Eq:Phi:dATI}
	\begin{aligned}
		\relaxket{\Tilde{\Phi}_{\text{d}}(v_{\text{f}},t)}
			&\propto \sum_{n=0}^{\infty}
				\int \dd \boldsymbol{\theta}\  
					x_1 e^{-i S_{\text{QO}}( \boldsymbol{\theta})/\hbar}
					 \dfrac{\alpha(\boldsymbol{\theta})^n}{\sqrt{n!}}
					 \\&\quad
					 \times
						 \big[
						 	E_{\text{cl}}(t_1)
						 	- E_{\alpha(\boldsymbol{\theta})}(\xi,t_1)
						 	+ \hat{E}_L(\xi,t_1)
						 \big]
						\ket{n}.
	\end{aligned}
\end{equation}
Here, for clarity, we denote $\boldsymbol{\theta} \equiv (t_1,x_2,v,x_1)$, and $E_{\alpha(\boldsymbol{\theta})}(\xi,t_1) = \langle \alpha(\boldsymbol{\theta})\vert \hat{E}_L(\xi,t)\vert\alpha(\boldsymbol{\theta})\rangle$, the latter obtained by moving the displacement operator in front of the vacuum state. The term $S_{\text{QO}}(\boldsymbol{\theta})$ denotes a modified semiclassical action incorporating quantum optical corrections, given by (see Appendix~\ref{Sec:App:SPA})
\begin{equation}
	\begin{aligned}
	S_{\text{QO}}(\boldsymbol{\theta})
		&= 
			S_{\text{cl}}(v,t,t_1)
			- x_2
				\big(
					v - v_{\text{f}}
				\big)
			- i\hbar \dfrac{\abs{\alpha(\boldsymbol{\theta})}^2}{2}
			\\&\quad
			- \dfrac{i}{2}
				x_1
					\big[
						i2\big(v_{\text{f}} + \mathsf{e}A_{\text{cl}}(t_1)\big)
						+ x_1 \alpha
					\big].
	\end{aligned}
\end{equation}
In the limit $\epsilon \to 0$, we have $\alpha(\boldsymbol{\theta}) \to 0$, and the integration variables can be regrouped to yield Dirac delta functions that recover the semiclassical dATI probability amplitude~\cite{lewenstein_rings_1995,amini_symphony_2019}.

Writing the final state as in Eq.~\eqref{Eq:Phi:dATI} proves beneficial, as all integrals concern only the probability amplitudes in the state. Since these involve highly oscillatory integrals, the saddle-point approximation becomes a particularly suitable tool for their evaluation~\cite{lewenstein_theory_1994,rivera-dean_role_2024}. This method allows one to express the integrals as a sum over carefully selected points---namely, the saddle-points of $S_{\text{QO}}(\boldsymbol{\theta})$, satisfying $\nabla_{\boldsymbol{\theta}} S_{\text{QO}}(\boldsymbol{\theta}) = \boldsymbol{0}$.~We find, however, that the explicit form of the saddle-point equations generally differs depending on the type of squeezing considered (see Appendix~\ref{Sec:App:SPA}). Nonetheless, both phase and amplitude squeezing lead the same structure for the saddle-point equation governing ionization
\begin{equation}
		\dfrac{\big[v+\mathsf{e}A_{\text{cl}}(t_1)\big]^2}{2}
			+ I_p
			- \mathsf{e}x_1E_{\text{cl}}(t_1)
			- \dfrac{i\hbar}{2} \pdv{\abs{\alpha(\boldsymbol{\theta})}^2}{t_1}
		= 0,
\end{equation}
where we see that in addition to the semiclassical saddle-point equation contributions, we identify two extra terms that effectively modify the ionization potential experienced by the electron.~The first, $\mathsf{e}x_1 E_{\text{cl}}(t_1)$, arises as an indirect consequence of light-matter entanglement at the moment of ionization [Fig.~\ref{Fig:purity:ion}]. The second, involving $\alpha(\boldsymbol{\theta})$, reflects a modification of the field amplitude induced by the electron's own dynamics.

Of the two additional contributions, the most intriguing one is that arising from the quantum optical displacement $\alpha(\boldsymbol{\theta})$.~As seen in Eq.~\eqref{Eq:def:alpha}, this quantity explicitly depends on the final measurement time $t$, meaning that the solution to the saddle-point equations is inherently influenced by the time at which the photoelectron is detected.~This explicit time dependence emerges as a direct consequence of entanglement between the electron and the quantized field.~In semiclassical treatments, the field is described by a mean-field approximation, where the quantum optical state is effectively assumed to remain unchanged by its interaction with the electron. As a result, the saddle-point equations derived in that context depend only on instantaneous field conditions and are independent on the measurement time.

By contrast, when adopting a fully quantum-optical description---where the field and electron interact bidirectionally---the field retains memory of its prior interactions with the electron, and vice versa. This is encoded in the displacement $\alpha(\boldsymbol{\theta})$: the quantum state of the field is perturbed by the electronic motion, which in turn influences the electron's trajectories.~The resulting entanglement between both subsystems introduces a history-dependent feedback mechanism:~the field's state at time $t$ reflects the electron's dynamics over the time interval $[t_1,t]$.~As a consequence, the obtained saddle-point equations become explicitly dependent on the final measurement time $t$.

\begin{figure}
	\centering
	\includegraphics[width = 1\columnwidth]{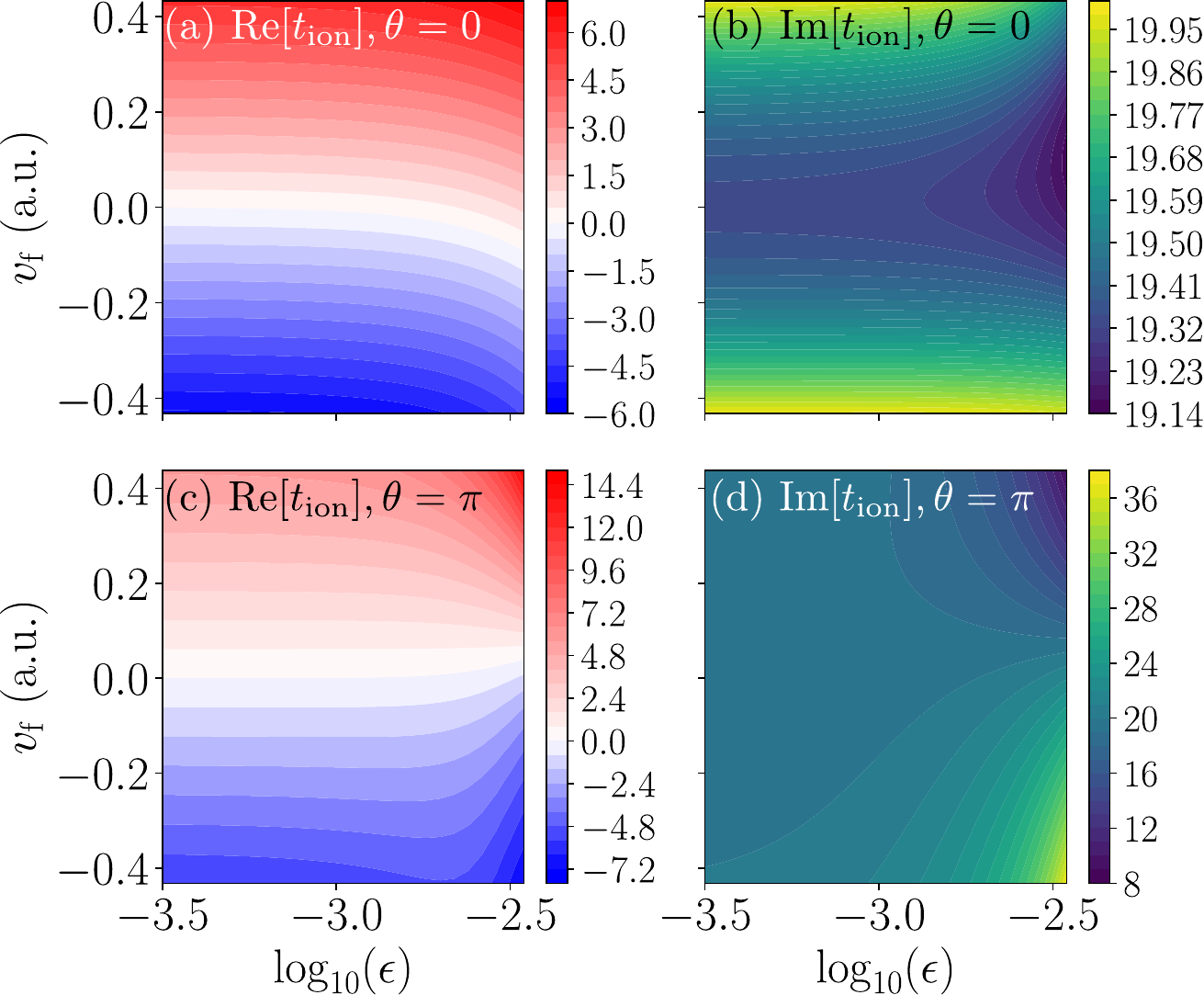}
	\caption{Real and imaginary parts of the ionization time as a function of $\epsilon = e^rg(\omega_L)$.~Panels (a) and (b) correspond to phase squeezing ($\theta = 0$), while panels (c) and (d) correspond to amplitude squeezing ($\theta = \pi$).~In all cases, we set $E_0 = 0.053$ a.u., $\omega_L = 0.057$ a.u., and the final measurement time to $t = 2\pi/\omega_L$.}
	\label{Fig:Ion:times}
\end{figure}

\begin{figure*}
	\centering
	\includegraphics[width=1\textwidth]{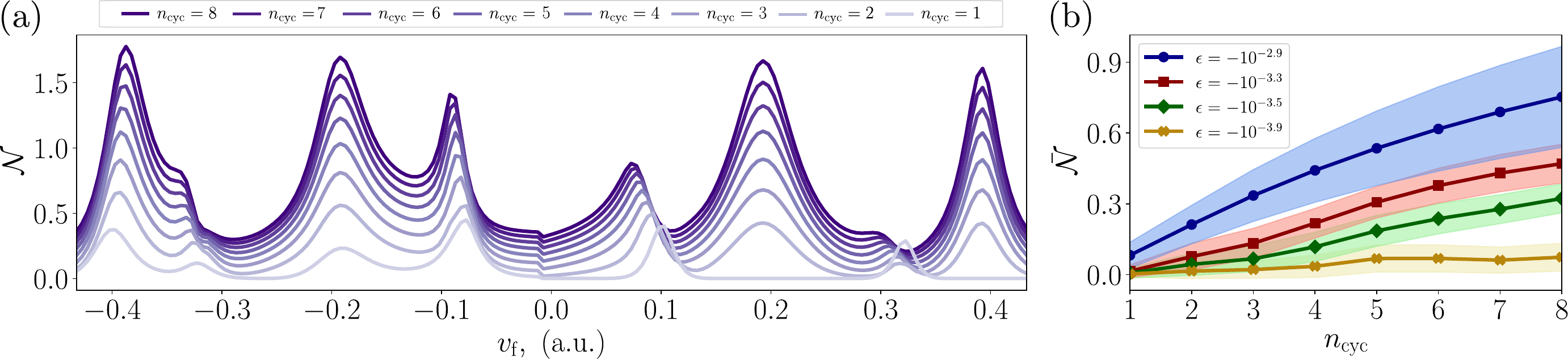}
	\caption{Behavior of the negative volume as a function of $\epsilon$ and $n_{\text{cyc}}$. Panel (a) shows the negative volume as a function of the final photoelectron kinetic momentum for different values of $n_{\text{cyc}}$, while $\epsilon = 10^{-2.9}$. Panel (b) shows the average value of $\mathcal{N}$ (solid curve) and its fluctuations (dashed region) as a function of $n_{\text{cyc}}$ and for different values of $\epsilon$.}
	\label{Fig:Negativity}
\end{figure*}

To illustrate how the type of squeezing modifies the ionization times, in Fig.~\ref{Fig:Ion:times} we display the real and imaginary parts of the ionization times within the interval $t_1 \in [-\pi/(2\omega_L),\pi/(2\omega_L)]$---shown in the left and right columns, respectively---for both phase and amplitude squeezing, displayed in the first and second rows respectively. As observed, for small amounts of squeezing, presented in terms of $\epsilon = e^{r}g(\omega_L)$, the ionization times coincide for both phase and amplitude squeezing, and are symmetric around $v_{\text{f}} = 0$ a.u., with $v_{\text{f}} < 0$ ($v_{\text{f}} > 0$) ionizing to the left (right) side of the field maximum (at $t=0$).

As the squeezing increases, this symmetry becomes increasingly broken, depending on the squeezing type.~For phase squeezing (panels (a) and (b)), the real parts of all ionization times shift towards earlier times, compensating for the stronger acceleration caused by the enhanced field fluctuations that follow.~These larger fluctuations also facilitate tunneling, resulting in overall reduced imaginary components of the ionization time. In contrast, for amplitude squeezing (panels (c) and (d)), the modifications are more symmetric.~The real part of the ionization time remains nearly symmetric around $v_{\text{f}} = 0$ a.u.---as expected since in those cases $\text{Re}[t_{\text{ion}}] = 0$ a.u., where the field fluctuations are minimal. For $v_{\text{f}} < 0$ ($v_{\text{f}} > 0$), the ionization times shift to the left (right) of the field maximum, as these occur in regions where the field fluctuations are stronger than at the maximum.~The imaginary part, however, reveals an asymmetry: ionization events with $v_{\text{f}} > 0$ experience a reduction in $\text{Im}[t_{\text{ion}}]$, indicating facilitated tunneling, whereas for $v_{\text{f}} > 0$, $\text{Im}[t_{\text{ion}}]$ increases, reflecting a higher tunneling barrier.

\subsection{Quantifying non-classical properties}
After applying the saddle-point approximation to compute the probability amplitudes in Eq.~\eqref{Eq:Phi:dATI}, we can rewrite the quantum state, up to a normalization factor, as
\begin{equation}\label{Eq:Phi:dATI:SPA}
	\begin{aligned}
	\relaxket{\Tilde{\Phi}_{\text{d}}(v_{\text{f}},t)}
		&= \sum_{\boldsymbol{\theta}_s}
				G(\boldsymbol{\theta}_s)
					\hat{D}_L
						\big(
							\alpha(\boldsymbol{\theta}_s)
						\big)
					\\&\hspace{1cm}\times
					\big[
						E_{\text{cl}}(t_{1,s})
						+ \hat{E}_L(\xi,t_{1,s})
					\big]
					\ket{0},
	\end{aligned}
\end{equation}
where $G(\boldsymbol{\theta}_s)$ is a complex-valued prefactor that arises from evaluating the integrand at the saddle-points, combined with additional weights that account for the contribution of each saddle-point~\cite{lewenstein_theory_1994,olga_simpleman,nayak_saddle_2019} (see Appendix~\ref{Sec:App:SPA} for more details). From this expression, we observe that the resulting state is generally non-classical, as it comprises a superposition of coherent states with different amplitudes $\alpha(\boldsymbol{\theta}_s)$, as well as displaced Fock states, arising from the term $\hat{D}(\alpha(\boldsymbol{\theta}_s))\hat{E}_L(\xi,t_{1,s}) \ket{0} \propto \hat{D}(\alpha(\boldsymbol{\theta}_s)) \ket{1}$. Importantly, the degree of non-classicality increases with both $\epsilon$ and the number of saddle-point solutions---the former by making each of the $\alpha(\boldsymbol{\theta}_s)$ more distinct, and the latter by adding more components to the total superposition.~Because the number of saddle-points is directly connected to the number of critical points in the field, we anticipate a corresponding dependence of the state's non-classical character on the number of optical cycles $n_{\text{cyc}}$.

In this subsection, our main objective is to characterize the potential non-classical properties of the quantum optical, and their dependence on both $n_{\text{cyc}}$ and $\epsilon$.~To properly analyze these features, we consider the quantum optical state in its original frame of reference, i.e., $\relaxket{\Phi_{\text{d}}(v_{\text{f}},t)} = \hat{D}_L(\alpha_L)\hat{S}(\xi)\relaxket{\Tilde{\Phi}_{\text{d}}(v_{\text{f}},t)}$.~However, this transformation complicates the numerical evaluation:~while displacement operators do not hinder the key properties of ~important non-classical witnesses such as the Wigner function or the covariance matrix, squeezing operations instead do modify them.~Here, the amount of squeezing considered is extremely large, requiring very high cutoffs in the Fock basis for accurate numerical implementation (see Appendix~\ref{Sec:App:Numerics} for more details).

To circumvent this issue, we use the negative volume $\mathcal{N}$, a measure of non-classicality defined as~\cite{kenfack_negativity_2004,walschaers_non-gaussian_2021}
\begin{equation}
	\mathcal{N}	
		= -1 + \int \dd x 
					\int \dd p\
						 \abs{W(x,p)},
\end{equation}
which quantifies the amount of negativity in the Wigner function $W(x,p)$---a hallmark of non-Gaussianity, and therefore of non-classical behavior~\cite{kenfack_negativity_2004}. Importantly for our purposes, this measure is invariant under Gaussian transformations~\cite{walschaers_non-gaussian_2021} such as displacement and squeezing operations (see Appendix~\ref{Sec:App:QO}). This invariance implies that computing the negative volume of $\relaxket{\Phi_{\text{d}}(v_{\text{f}},t)}$---in the original frame of references---is equivalent to computing it for $\relaxket{\Tilde{\Phi}_{\text{d}}(v_{\text{f}},t)}$---in the displaced and squeezed frame of reference---with the latter requiring significantly fewer numerical resources.

Figure~\ref{Fig:Negativity} presents the results of our analysis in the case of phase squeezing.~Panel (a) shows the negativity $\mathcal{N}$ as a function of the final kinetic momentum $v_{\text{f}}$ of the measured photoelectron for different numbers of optical cycles, with $\epsilon = 10^{-2.9}$ fixed.~Panel (b) displays the average of $\mathcal{N}$ (solid curves) and its fluctuations (dashed region), computed over the interval $v_{\text{f}} \in [-\sqrt{U_p},\sqrt{U_p}]$, as a function of $n_{\text{cyc}}$ for several values of $\epsilon$. As expected, we observe that both larger $n_{\text{cyc}}$ and $\epsilon$ result in more overall pronounced non-classical features, as captured by $\mathcal{N}$.

Interestingly, panel (a) reveals that the negativity exhibits a strong dependence on the final kinetic momentum $v_{\text{f}}$, with more prominent features appearing at specific values of $v_{\text{f}}$.~These features appear to be approximately symmetric around $v_{\text{f}} = 0$. In particular, we observe that the energy separation between the two most prominent peaks---located at $\abs{v_{\text{f}}} \approx 0.2$ a.u. and $\abs{v_{\text{f}}} \approx 0.4$ a.u.---is approximately $\Delta E_{v} = 0.056$ a.u., which is comparable to the spacing between ATI peaks observed in standard ATI spectra~\cite{paulus_plateau_1994,hansch_resonant_1997}.~This suggests that some of the non-classical features may be correlated with the ATI structure, although additional peaks that do not align strictly with ATI energies are also observed.~However, unlike standard ATI spectra, for large values of $n_{\text{cyc}}$ ($n_{\text{cyc}} > 1$), the negativity profile is not fully symmetric about $v_{\text{f}} = 0$; in fact, we observe a small discontinuity at $v_{\text{f}} \approx 0$. This asymmetry arises because the number of saddle-points differs between positive and negative $v_{\text{f}}$ in our calculations. In our setup, we fix the measurement time to $t= 2\pi n_{\text{cyc}}/\omega_L$, which implies that trajectories contributing to $v_{\text{f}} < 0$ include an additional ionization event just before the field maximum at $t$, whereas for $v_{\text{f}} >0$, the extra ionization event would occur after the maximum (see Fig.~\ref{Fig:Ion:times} for reference).

\subsection{Light-matter entanglement properties}
In Ref.~\cite{rivera-dean_light-matter_2022} it was shown that the electron's backaction on the quantum optical state can lead to the emergence of light-matter entanglement when considering coherent state drivers.~However, such entanglement features were found to be negligible in the the near-infrared regime ($\lambda_L \approx 800$ nm), and only mildly present at mid-infrared wavelengths ($\lambda_L \approx 2000$ nm). In this section, motivated by the emergence of prominent non-Gaussian features in the final quantum optical state---and their dependence on the the final electronic momentum---we investigate whether the introduction of squeezing features in the driver can enhance these correlations.

To address this question---and in contrast to Ref.~\cite{rivera-dean_light-matter_2022}, which focused on electrons propagating with a specific kinetic energy and in opposite directions---we characterize the properties of the driving field after performing the projective measurement $\hat{\Pi}_{\text{lim}} = \int^{v_{\text{lim}}}_{-v_{\text{lim}}} \dd v \dyad{v}$ on the electronic degrees of freedom. This operator describes a measurement that filters electron whose kinetic energy lies within the range $[0,v_{\text{lim}}^2/2]$, and together with its complement $\hat{\bar{\Pi}} = \mathbbm{1}-\hat{\Pi}_{\text{lim}}$, forms a complete measurement basis~\cite{NielsenBookCh1}. Accordingly, when $v_{\text{lim}} < \sqrt{4U_p}$, applying $\hat{\Pi}_{\text{lim}}$ to Eq.~\eqref{Eq:Total:state} yields, up to normalization,
\begin{equation}\label{Eq:post:meas:state}
	\begin{aligned}
	\hat{\rho}
		&= \tr[\hat{\Pi}_{\text{lim}} \relaxket{\Tilde{\Psi}(t)}\!\relaxbra{\Tilde{\Psi}(t)}]
		\approx \int_{-v_{\text{lim}}}^{v_{\text{lim}}} \!\!\!\dd v
			\relaxket{\Tilde{\Phi}_{\text{d}}(v,t)}
			\!\relaxbra{\Tilde{\Phi}_{\text{d}}(v,t)},
	\end{aligned}
\end{equation}
where the ground state and HATI components in Eq.~\eqref{Eq:Total:state} vanish, as electrons with the considered energy range are predominantly produced through dATI processes.~The resulting state is generally mixed, and since the original state before the measurement was pure, the degree of mixedness reflects the amount of entanglement that was present in the original pure light-matter state.~To quantify this, we use the linear entropy $S_{\text{lin}}(\hat{\rho}) = 1 - \tr(\hat{\rho}^2)$~\cite{agarwal_quantitative_2005,berrada_beam_2013}, a particularly suitable entanglement measure for systems with infinite-dimension Hilbert spaces.

\begin{figure}
	\centering
	\includegraphics[width=1\columnwidth]{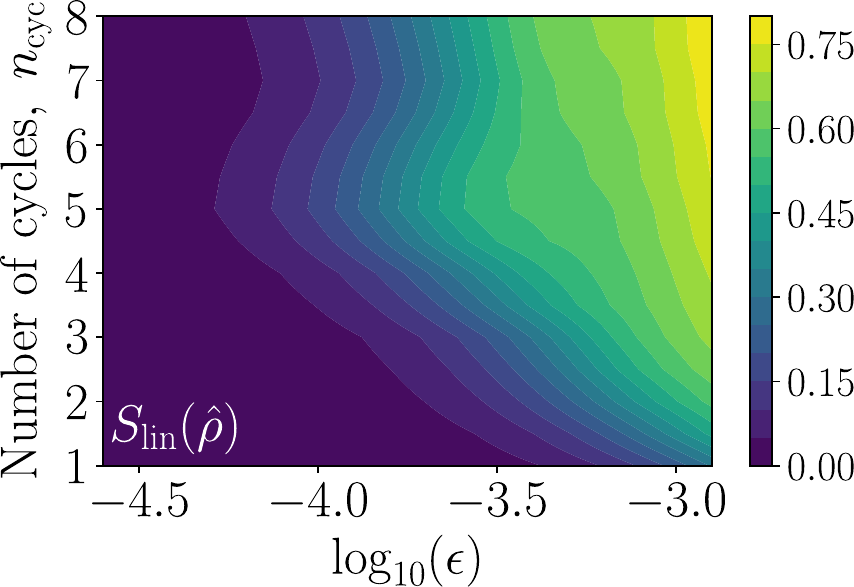}
	\caption{Linear entropy as a function of the amount of squeezing, quantified through $\epsilon$, and the number of field cycles $n_{\text{cyc}}$.}
	\label{Fig:Lin:entropy}
\end{figure}

Figure~\ref{Fig:Lin:entropy} shows the linear entropy of the post-selected state in Eq.~\eqref{Eq:post:meas:state} for the case of a phase-squeezed driver, plotted as a function of both the number of cycles in the driving field---which determines the number of ionization events between the initial and measurement times---and the squeezing amplitude $\epsilon$.~As expected from Ref.~\cite{rivera-dean_light-matter_2022}, in the limit of vanishing squeezing, the linear entropy tends to zero, indicating negligible light-matter entanglement.~As $\epsilon$ increases, entanglement features become more pronounced, with the effect amplified by a greater number of ionization events, that is, larger $n_{\text{cyc}}$.~This results in significant entanglement features for sufficiently large values of both parameters, reaching a maximum of $\text{max}_{\epsilon,n_{\text{cyc}}}[S_{\text{lin}}(\hat{\rho})] \approx 0.79$ for the range of parameters considered here.~Interestingly, for a fixed value of $\epsilon$, we observe the emergence of a saturation point beyond which increasing the number of cycles no longer leads to further entanglement growth.~This saturation occurs around $n_{\text{cyc}} = 5$ for $\log_{10}(\epsilon)\in[-4.25-3.4]$.~Notably, for squeezing amplitudes compatible with current experimental capabilities---namely $\epsilon \approx 10^{-3.5}$ a.u.~\cite{rasputnyi_high_2024,tzur_measuring_2025}---we find a substantial degree of entanglement, with $S_{\text{lin}}(\hat{\rho}) \approx 0.5$, provided the driving field spans at least ten optical cycles.

\section{CONCLUSIONS}
In this work, we have studied the process of direct ATI driven by intense squeezed light, and explored its impact on the non-classical properties of the joint electron-light state. We have shown that the presence of squeezing significantly enhances the light-matter coupling, rendering the mutual backaction between light and matter an essential effect to consider.~As a consequence of this enhanced interaction, the non-classical properties of the resulting state are markedly modified compared to the scenario where a classical coherent state driver is used~\cite{rivera-dean_light-matter_2022}.

Specifically, we found that the mutual backaction modifies the ionization events of the electron [Fig.~\ref{Fig:Ion:times}], as well as the entanglement properties of the joint state, both immediately after ionization [Fig.~\ref{Fig:purity:ion}], and at the final measurement time [Fig.~\ref{Fig:Lin:entropy}]. These effects amplify as the amount of squeezing in the driving field increases.~For the latter, we also observed that the number of ionization events occurring between the initial and final times---ultimately set by the number of optical cycles---strongly influences the degree of entanglement. Additionally, we showed that projecting the electronic state onto specific final momenta can result in non-Gaussian quantum optical states of the driving field [Fig.~\ref{Fig:Negativity}]. Remarkably, all these effects are predicted to be observable under squeezing levels achievable with current state-of-the-art experimental capabilities~\cite{manceau_indefinite-mean_2019,rasputnyi_high_2024,tzur_measuring_2025,lemieux_photon_2024}.

Finally, although this work has focused primarily on describing direct ATI processes, our formalism readily allows for extensions to high-order ATI events by explicitly incorporating the first-order perturbative contributions to Eq.~\eqref{Eq:perb:theory:ATI}~\cite{lewenstein_rings_1995,lohr_above-threshold_1997}, or to below-threshold non-sequential double ionization, which can be described as time-ordered ATI-like processes~\cite{maxwell_quantum_2015,maxwell_controlling_2016}.~More intriguingly, given the $S$-matrix-like structure adopted in our approach~\cite{lohr_above-threshold_1997}, it would be particularly valuable to explore how path-integral techniques~\cite{milosevic_phase_2013} might be integrated with this quantum optical framework.~Such an integration could enable the development of Coulomb-distorted SFA approaches~\cite{lai_influence_2015,maxwell_controlling_2016,lai_near-threshold_2017,maxwell_coulomb-corrected_2017,maxwell_analytic_2018,maxwell_coulomb-free_2018} within a fully quantized field description, potentially yielding exact analytical expressions for expectation values of the propagator in Eq.~\eqref{Eq:Prop:Cont}.

\section*{ACKNOWLEDGEMENTS}
J.R.-D.~acknowledges insightful discussions with Emilio Pisanty during the development of Ref.~\cite{rivera_dean_quantum-optical_2019}, which helped inspire the formalism presented in this work.~Valuable conversations with Thomas Rook and Gurpahul Singh are also gratefully acknowledged. C.F.M.F.~would like to thank ICFO for its kind hospitality.

ICFO-QOT group acknowledges support from:
European Research Council AdG NOQIA; MCIN/AEI (PGC2018-0910.13039/501100011033,  CEX2019-000910-S/10.13039/501100011033, Plan National FIDEUA PID2019-106901GB-I00, Plan National STAMEENA PID2022-139099NB, I00, project funded by MCIN/AEI/10.13039/501100011033 and by the “European Union NextGenerationEU/PRTR" (PRTR-C17.I1), FPI); QUANTERA DYNAMITE PCI2022-132919, QuantERA II Programme co-funded by European Union’s Horizon 2020 program under Grant Agreement No 101017733; Ministry for Digital Transformation and of Civil Service of the Spanish Government through the QUANTUM ENIA project call - Quantum Spain project, and by the European Union through the Recovery, Transformation and Resilience Plan - NextGenerationEU within the framework of the Digital Spain 2026 Agenda; Fundació Cellex; Fundació Mir-Puig; Generalitat de Catalunya (European Social Fund FEDER and CERCA program; Barcelona Supercomputing Center MareNostrum (FI-2023-3-0024); Funded by the European Union. Views and opinions expressed are however those of the author(s) only and do not necessarily reflect those of the European Union, European Commission, European Climate, Infrastructure and Environment Executive Agency (CINEA), or any other granting authority.  Neither the European Union nor any granting authority can be held responsible for them (HORIZON-CL4-2022-QUANTUM-02-SGA  PASQuanS2.1, 101113690, EU Horizon 2020 FET-OPEN OPTOlogic, Grant No 899794, QU-ATTO, 101168628),  EU Horizon Europe Program (This project has received funding from the European Union’s Horizon Europe research and innovation program under grant agreement No 101080086 NeQSTGrant Agreement 101080086 — NeQST); ICFO Internal ``QuantumGaudi'' project;

P. Stammer acknowledges support from: The European Union’s Horizon 2020 research and innovation programme under the Marie Skłodowska-Curie grant agreement No 847517.

\bibliography{References.bib}{}

\newpage
\clearpage
\onecolumngrid
\appendix

\begin{center}
    {\large \textbf{\textsc{Appendix}}}
\end{center}

\section{ANALYSIS OF THE LIGHT-MATTER INTERACTION DYNAMICS}
\subsection{Light-matter interaction Hamiltonian}\label{Sec:App:Hamiltonian}
In this work, we aim to solve the light-atom interaction dynamics---within the single-active electron and dipole approximation---governed by the following time-dependent Schrödinger equation (TDSE)
\begin{equation}\label{Eq:App:init:SE}
	i\hbar \pdv{\ket{\Psi(t)}}{t}
		= \big[
					\hat{H}_{\text{at}}
					+ \mathsf{e}\hat{r}\hat{E}
					+ \hat{H}_{\text{field}}
			\big]\ket{\Psi(t)},
\end{equation}
subject to the initial condition $\ket{\Psi(t_0)} = \ket{\text{g}} \otimes \ket{\xi, \alpha} \bigotimes_{q\neq 1}\ket{0_q}$, where $\ket{\text{g}}$ denotes the atomic ground state, and $\ket{\xi,\alpha} \equiv \hat{D}_q(\alpha) \hat{S}_q(\xi)\ket{0}$ is a displaced squeezed vacuum (DSV) state in the $q$th harmonic order (in the following we denote $q=1\equiv L$). In Eq.~\eqref{Eq:App:init:SE}, $\hat{H}_{\text{at}}$ represents the atomic Hamiltonian, $\mathsf{e}\hat{r}$ the dipole operator, $\hat{E}(t)$ the electric field operator and $\hat{H}_{\text{field}}$ the free-field Hamiltonian.

When transitioning to the interaction picture with respect to the free-field Hamiltonian, the electric field operator becomes time-dependent and takes the form
\begin{equation}
	\hat{E}(t)
		 = \sum_{q=1} \hat{E}_q(t)
		 = -i \sum_{q=1}g(\omega_q)
		 			\big[
		 				\hat{a}_q e^{-i\omega_q t}
		 				- \hat{a}^\dagger_q e^{i\omega_q t}
		 			\big],			
\end{equation} 
where $g(\omega_q) \equiv \sqrt{\hbar \omega_q/(2\epsilon_0 V)}$ is a mode-dependent prefactor that arises from the expansion of the electric field operator in terms of the creation and annihiliation operators $\hat{a}^\dagger_q$ and $\hat{a}_q$, which respectively create and destroy energy quanta in the $q$th harmonic mode.~In our context, $g(\omega_q)$ represents the coupling between the $q$th harmonic mode with matter. This interaction picture is introduced by defining the time-dependent state as $\ket{\Psi(t)} = e^{-i \hat{H}_{\text{field}} t/\hbar}\lvert\check{\Psi}(t)\rangle$, where $\relaxket{\check{\Psi}(t)}$ evolves according to
\begin{equation}\label{Eq:App:Hfield:SE}
	i\hbar \pdv{\lvert\check{\Psi}(t)\rangle}{t}
		= \big[
				\hat{H}_{\text{at}}
				+ \mathsf{e}\hat{r}\hat{E}(t)
			\big]\lvert\check{\Psi}(t)\rangle.
\end{equation}

Next, we define $\relaxket{\check{\Psi}(t)}= \hat{D}_L(\alpha)\relaxket{\bar{\Psi}(t)}$, which corresponds to moving into a frame displaced by the initial coherent state amplitude of the driving field mode. In this displaced frame, the initial state takes the form $\relaxket{\bar{\Psi}(t_0)} = \ket{\text{g}} \otimes \hat{S}_{L}(\xi)\relaxket{\bar{0}}$,  where $\relaxket{\bar{0}} \equiv \bigotimes_{q} \ket{0_q}$ denotes the vacuum state across all modes. Making use of the identity $\hat{D}^\dagger(\alpha)\hat{a}\hat{D}(\alpha) = \hat{a} + \alpha$, the TDSE above transforms to
\begin{equation}\label{Eq:App:Disp:SE}
	i\hbar \pdv{\relaxket{\bar{\Psi}(t)}}{t}
		= \big[
				\hat{H}_{\text{at}}
				+ \mathsf{e} \hat{r}E_{\text{cl}}(t)
						+ \mathsf{e} \hat{r}\hat{E}(t)
			\big]\relaxket{\bar{\Psi}(t)},
\end{equation}
with $E_{\text{cl}}(t) = \relaxbra{\alpha}\hat{E}_1(t)\relaxket{\alpha} = -i g(\omega_L)[\alpha e^{-i\omega_L t} - \alpha^* e^{i\omega_L t}]$. Within this framework, analogies with semiclassical strong-field physics become more transparent, allowing us to leverage established analytical techniques~\cite{lewenstein_theory_1994,amini_symphony_2019}.

Finally, we perform the transformation $\relaxket{\bar{\Psi}(t)} = \hat{S}_L(\xi)\relaxket{\Tilde{\Psi}(t)}$, which brings the initial quantum optical state to the vacuum across all modes, i.e., $\relaxket{\Tilde{\Psi}(t_0)} = \ket{\text{g}}\otimes \relaxket{\bar{0}}$.~Using the identity $\hat{S}^\dagger(\alpha)\hat{a}\hat{S}(\alpha) = \hat{a}\cosh(r)-\hat{a}^\dagger e^{i\theta}\sinh(r)$, where $\xi = re^{i\theta}$ ($r>0$), the component of the electric field operator acting on the driving field mode transforms as
\begin{equation}\label{Eq:App:Squeez:E}
	\begin{aligned}
	\hat{E}_L(\xi,t) &\equiv
		\hat{S}_L^\dagger(\xi)\hat{E}_L(t)\hat{S}_L(\xi)
			\\&= -i g(\omega_L)
				\Big[
					\big(
						\!\cosh(r)e^{-i\omega_L t} 
						+ e^{i(\omega_L t-\theta)}\sinh(r)
					\big) \hat{a}_L
					-
					\big(
						\!\cosh(r)e^{i\omega_L t} 
						+ e^{-i(\omega_L t-\theta)}\sinh(r)
					\big) \hat{a}^\dagger_L
				\Big]
			\\& = -i
				\big[
					f(\xi,t)\hat{a}_L
					- f^*(\xi,t)\hat{a}^\dagger_L
				\big],
	\end{aligned}
\end{equation}
where we define $f(\xi,t) \equiv g(\omega_L)[\cosh(r)e^{-i\omega_Lt} + e^{i(\omega_L t - \theta)}\sinh(r)]$.~This expression shows explicitly how squeezing modifies and enhances the effective light-matter coupling.~Under this transformation, the TDSE becomes
\begin{equation}\label{Eq:App:Squ:SE}
	i\hbar
		\pdv{\relaxket{\tilde{\Psi}(t)}}{t}
			= \big[
					\hat{H}_{\text{at}}
					+ \mathsf{e}\hat{r}E_{\text{cl}}(t)
					+ \mathsf{e}\hat{r}\hat{E}_{L}(\xi,t)
					+ \mathsf{e}\hat{r}\hat{E}_{\text{uv}}(t)
				\big]\relaxket{\Tilde{\Psi}(t)},
\end{equation}
where we denote $\hat{E}_{\text{uv}}(t) = \sum_{q>1}\hat{E}_q(t)$.~Equation \eqref{Eq:App:Squ:SE} defines the central dynamical equation used in this work.

\subsection{Solving the time-dependent Schrödinger equation}\label{Sec:App:TDSE}
To solve Eq.~\eqref{Eq:App:Squ:SE}, we adopt a strategy inspired by the approach in Ref.~\cite{rivera_dean_quantum-optical_2019}, which aims to construct the time-evolution operator $\hat{U}(t,t_0)$ that propagates the initial state according to $\relaxket{\Tilde{\Psi}(t_0)} \to \relaxket{\Tilde{\Psi}(t)} = \hat{U}(t,t_0)\relaxket{\Tilde{\Psi}(t_0)}$. To outline the method in general terms, let us consider a Hamiltonian of the form $\hat{H}(t) = \hat{H}_0(t) + \hat{V}(t)$, where $\hat{H}_0(t)$ is a Hamiltonian whose evolution can be handled analytically, and $\hat{V}(t)$ is a time-dependent interaction. In this setting, the time-evolution operator satisfies the differential equation
\begin{equation}\label{Eq:App:TDSE:U}
	i \hbar \pdv{\hat{U}(t)}{t}
		= \big[
				\hat{H}_0(t) + \hat{V}(t)
			\big]\hat{U}(t).
\end{equation}

A solution to the differential equation above can always be written in the form~\cite{smirnova_anatomy_2007}
\begin{equation}\label{Eq:App:general:Dyson}
	\hat{U}(t,t_0)
		= \hat{U}_0(t,t_0)
			- \dfrac{i}{\hbar}
				\int^t_{t_0} \dd t_1
					\hat{U}(t,t_1)
						\hat{V}(t_1)
							\hat{U}_0(t_1,t_0),
\end{equation}
where $\hat{U}_0(t,t_0)$ describes the evolution under the unperturbed Hamiltonian $\hat{H}_0(t)$, i.e., it satisfies $i\hbar \partial{\hat{U}_0(t)}/{t} = \hat{H}_0(t)\hat{U}_0(t)$. Equation \eqref{Eq:App:general:Dyson} serves as the basis for a recursive formulation, where $\hat{U}(t,t_1)$ accounts for the full dynamics governed by $\hat{H}_0(t) + \hat{V}(t)$.~The key idea used in Refs.~\cite{lohr_above-threshold_1997,rivera_dean_quantum-optical_2019} is to adopt different different partitions of the total Hamiltonian at each recursive step---effectively reassigning what is considered the ``free'' part $\hat{H}_0(t)$ and what to remains in the interaction $\hat{V}(t)$. In the case of Eq.~\eqref{Eq:App:Squ:SE}, we can naturally distinguish two contributions to the interaction
\begin{equation}
	\hat{H}_0
		 = \hat{H}_{\text{at}}
	\quad
	\hat{V}_L(t)
		 = \mathsf{e}\hat{r}E_{\text{cl}}(t)
		 	 + \mathsf{e}\hat{r}\hat{E}_L(\xi,t),
	\quad
	\hat{V}_{\text{uv}}(t)
		= \mathsf{e}\hat{r}\hat{E}_{\text{uv}}(t),
\end{equation}
and initially choose the decomposition $\hat{H}_0 = \hat{H}_{\text{at}}$ and $\hat{V}(t) = \hat{V}_L(t) + \hat{V}_{\text{uv}}(t)$. With this partition, the Dyson expansion \eqref{Eq:App:general:Dyson} reads
\begin{equation}\label{Eq:App:first:partition}
	\hat{U}(t,t_0)
		= \hat{U}_{\text{at}}(t,t_0)
			- \dfrac{i}{\hbar}
				\int^t_{t_0} \dd t_1
					\hat{U}(t,t_1)
							\big[\hat{V}_L(t_1) + \hat{V}_{\text{uv}}(t_1)\big]
							\hat{U}_{\text{at}}(t_1,t_0),
\end{equation}
where $\hat{U}_{\text{at}}(t)$ denotes the propagator generated by the atomic Hamiltonian alone, i.e., it solves $i\hbar \partial{\hat{U}_{\text{at}}(t)}/{t} = \hat{H}_{\text{at}}\hat{U}_{\text{at}}(t)$.

For the second iteration, we redefine the partition as $\hat{H}_0(t) = \hat{H}_{\text{at}} + \hat{V}_L(t)$ and $\hat{V}(t) = \hat{V}_{\text{uv}}(t)$.~This yields an equally valid solution to Eq.~\eqref{Eq:App:TDSE:U}, written as
\begin{equation}\label{Eq:App:second:partition}
	\hat{U}(t,t_0)
	= \hat{U}_{L}(t,t_0)
		- \dfrac{i}{\hbar}
			\int^t_{t_0} \dd t_1
				\hat{U}(t,t_1)
						 \hat{V}_{\text{uv}}(t_1)
						\hat{U}_{L}(t_1,t_0),
\end{equation}
where $\hat{U}_{L}(t)$ denotes the evolution operator governed by the atomic Hamiltonian plus the interaction with the driving field, satisfying $i\hbar \partial{\hat{U}_{L}(t)}/{t} = [\hat{H}_{\text{at}} + \hat{V}_L(t)]\hat{U}_{L}(t)$. This operator involves only the degrees of freedom associated with the driving optical mode.~Inserting Eq.~\eqref{Eq:App:second:partition} into Eq.~\eqref{Eq:App:first:partition}, we obtain
\begin{equation}\label{Eq:App:gen:Uop}
	\begin{aligned}
	\hat{U}(t,t_0)
		&= \hat{U}_{\text{at}}(t,t_0)
			- \dfrac{i}{\hbar}
				\int^t_{t_0}\dd t_1
					\hat{U}_{L}(t,t_1)
						\big[
							\hat{V}_L(t_1) 
							+ \hat{V}_{\text{uv}}(t_1)
						\big]
							\hat{U}_{\text{at}}(t_1,t_0)
		\\&\quad
			- \dfrac{1}{\hbar^2}
				\int^{t}_{t_1} \dd t_2\int^{t}_{t_0} \dd t_1
					\hat{U}(t,t_2)
							\hat{V}_{\text{uv}}(t_2)
							\hat{U}_{L}(t_2,t_1)
								\big[
									\hat{V}_L(t_1) 
									+ \hat{V}_{\text{uv}}(t_1)
								\big]
									\hat{U}_{\text{at}}(t_1,t_0),
	\end{aligned}
\end{equation}
where retaining just the first two terms is already sufficient to describe ATI events, while the remaining contributions account for more complex processes such as HHG or UV-driven excitations.~In particular, by reinserting Eq.~\eqref{Eq:App:first:partition} in the expression above and neglecting higher-order terms, we arrive at
\begin{equation}\label{Eq:App:gen:Uop:approx}
	\begin{aligned}
		\hat{U}(t,t_0)
		&\approx \hat{U}_{\text{at}}(t,t_0)
				- \dfrac{i}{\hbar}
						\int^t_{t_0}\dd t_1
							\hat{U}_{L}(t,t_1)
							\big[
								\hat{V}_L(t_1) 
							+ \hat{V}_{\text{uv}}(t_1)
							\big]
							\hat{U}_{\text{at}}(t_1,t_0)
			\\&\quad
				- \dfrac{1}{\hbar^2}
						\int^{t}_{t_1} \dd t_2\int^{t}_{t_0} \dd t_1
							\hat{U}_{\text{at}}(t,t_2)
								\hat{V}_{\text{uv}}(t_2)
								\hat{U}_{L}(t_2,t_1)
								\big[
									\hat{V}_L(t_1) 
									+ \hat{V}_{\text{uv}}(t_1)
								\big]
									\hat{U}_{\text{at}}(t_1,t_0),
	\end{aligned}
\end{equation}

\subsection{The Strong-Field Approximation and the direct ATI contribution}\label{Sec:App:SFA:ATI}
To further advance our analysis and derive quasianalytical expressions for the quantum state of the joint light-matter system, we rely on the Strong-Field Approximation (SFA)~\cite{lewenstein_theory_1994,amini_symphony_2019}. Broadly speaking, the SFA encompasses a remarkably successful set of methods for simplifying the treatment of strongly driven light-matter interactions~\cite{amini_symphony_2019,armstrong_dialogue_2021}. In its standard formulation, two key assumptions are made:
\begin{enumerate}[(a)]
	\item The strong laser field does not couple to any bound state other than the ground state $\ket{\text{g}}$, so that only the ground state and the continuum states $\{\ket{k}\}$ are included in the dynamics.
	\item Once in the continuum, the electron is effectively treated as a free particle driven by the external electric field, with the Coulomb potential acting only as a perturbative correction to its motion.
\end{enumerate}

Assumption (a) allows us to define an SFA-version of the identity operator for the atomic Hilbert space as
\begin{equation}
	\mathbbm{1}
		 = \dyad{\text{g}}
		 	+ \int \dd k \dyad{k},
\end{equation}
which, when inserted between the time-evolution operator and a $\hat{V}(t)$-like interaction term in Eq.~\eqref{Eq:App:gen:Uop:approx} yields
\begin{align}
	\relaxket{\Tilde{\Psi}(t)}
		&= \hat{U}(t,t_0)\ket{\text{g}}\otimes \relaxket{\bar{0}}\nonumber
		\\&
		\approx \hat{U}_{\text{at}}(t)
				\ket{\text{g}} \otimes \relaxket{\bar{0}}
			- \dfrac{i}{\hbar}
				\int^t_{t_0} \dd t_1 \int \dd k\
					\hat{U}_{L}(t,t_1)
						\relaxbra{k}
							\big[
								\hat{V}_L(t_1) + \hat{V}_{\text{uv}}(t_1)
							\big]\relaxket{\text{g}}
						e^{i I_p (t_1-t_0)/\hbar}
							\ket{k}\otimes \ket{\bar{0}}\label{Eq:App:ATI:contrib}
		\\& \quad 
			-\dfrac{1}{\hbar^2} \int^{t}_{t_1}\! \dd t_2 \!\int^{t}_{t_0} \!\dd t_1 \!\int\! \dd k
			e^{i\frac{I_p}{\hbar}(t-t_2)}
				\relaxbra{\text{g}}
						\hat{V}_{\text{uv}}(t_2)
					 \hat{U}_L(t_2,t_1)\relaxket{k}
					\relaxbra{k}
					\big[
						\hat{V}_L(t_1) + \hat{V}_{\text{uv}}(t_1)
					\big]\relaxket{\text{g}}
					e^{i\frac{I_p}{\hbar}(t_1-t_0)}
					\ket{\text{g}}\otimes \relaxket{\bar{0}}\label{Eq:App:HHG:contrib}
			\\& \quad 
			-\dfrac{1}{\hbar^2} \int ^{t}_{t_1} \!\dd t_2\! \int^{t}_{t_0} \!\dd t_1\! \int\! \dd k_2\! \int\! \dd k_1
				\hat{U}_{\text{at}}(t,t_2)
					\relaxbra{k_2}
						\hat{V}_{\text{uv}}(t_2)
					 \hat{U}_L(t_2,t_1)\relaxket{k_1}
						\relaxbra{k_1}
						\big[
							\hat{V}_L(t_1) + \hat{V}_{\text{uv}}(t_1)
						\big]\relaxket{\text{g}}
						e^{i\frac{I_p}{\hbar}(t_1-t_0)}
						\ket{k_2}\otimes \relaxket{\bar{0}}\label{Eq:App:HATI:contrib}.
\end{align}
Here, we used $\mel{\text{g}}{\hat{r}}{\text{g}} = 0$ due to parity symmetry. Each of the resulting contributions---Eqs.~\eqref{Eq:App:ATI:contrib} to \eqref{Eq:App:HATI:contrib}---corresponds to a distinct strong-field mechanism, which can be individually identified by the structure of the interaction terms involved. Specifically:
\begin{itemize}
	\item In Eq.~\eqref{Eq:App:ATI:contrib}, we identify two distinct contributions.~The first term corresponds to the scenario where the electron remains in the ground state throughout the interaction, experiencing no coupling with the external electromagnetic field.~Consequently, the photonic state remains unchanged. The second term captures events in which the electron interacts with the field at time $t_1$---via either the driving field mode or the harmonic modes---resulting into a transition from the ground state $\ket{\text{g}}$ to a continuum state $\ket{k_1}$. From that point on, the light-matter system evolves under $\hat{U}_L(t,t_1)$ until the final time $t$. In the absence of the quantum optical interaction term $\mathsf{e}\hat{r}\hat{E}_L(t)$, $\hat{U}_L(t,t_1)$ describes the propagation of an electron driven by a classical field, including the effect of the atomic core---capturing both direct and high-order ATI. The inclusion of the quantum-optical coupling modifies this evolution, allowing for back-action on the field and vice versa, ultimately resulting in entanglement between light and matter~\cite{rivera-dean_light-matter_2022}.
	
	\item In contrast, Eq.~\eqref{Eq:App:HHG:contrib} features two-field induced transitions. The first occurs at time $t_1$, promoting the electron from the ground state to the continuum. The second, at time $t_2$, involves a recombination process in which the electron, after evolving through $\hat{U}_L(t_2,t_1)$, returns to the ground state while emitting a photon into the harmonic modes. Consequently, this process underlies HHG events.
	
	\item The final contribution, given in Eq.~\eqref{Eq:App:HATI:contrib}, is structurally similar to the HHG term, with the key distinction that the second transition occurs between two continuum states $\ket{k_2}$ and $\ket{k_1}$, and is mediated by interaction with the harmonic modes. Since the coupling to these modes is typically weak (proportional to $g(\omega_q)$ and unaffected by squeezing), this contribution constitutes a higher-order correction to Eq.~\eqref{Eq:App:ATI:contrib}. 
\end{itemize}

In this work, our main focus lies on the ATI contribution, namely,
\begin{equation}\label{Eq:App:ATI}
	\relaxket{\tilde{\Psi}_{\text{ATI}}(t)}
		= - \dfrac{i}{\hbar}
				\int^t_{t_0} \dd t_1
					\int \dd k \
						\hat{U}_L(t,t_1)
							\bra{k}
								\big[
									\hat{V}_L(t_1)
									+ \hat{V}_{\text{uv}}(t_1)
								\big]
							\ket{\text{g}}
							e^{iI_p(t_1-t_0)/\hbar}
							\ket{k} \otimes \ket{\bar{0}},
\end{equation}
which, in contrast to semiclassical analyses, explicitly incorporates the coupling between the quantum optical modes and the matter degrees of freedom. As mentioned earlier, and in contrast to approaches where the driving field lies in a coherent state~\cite{rivera-dean_light-matter_2022,stammer_quantum_2023}, this coupling is further enhanced by the presence of squeezing features in the driving field. It is important to emphasize that this expression is fully general: the influence of the atomic potential is encoded within $\hat{U}_L(t)$, allowing the expression to account for both direct and high-order ATI processes. The operator $\hat{U}_L(t)$ satisfies
\begin{equation}\label{Eq:App:UL:diff:eq}
	i\hbar \pdv{\hat{U}_L(t)}{t}
		= \Big[
				\dfrac{\hat{p}^2}{2m}
				+ V_{\text{at}}(\hat{r})
				+ \mathsf{e}\hat{r}
					\big(
						E_{\text{cl}}(t)
						+ \hat{E}_L(t)
					\big)
			\Big]\hat{U}_L(t),
\end{equation}
and, when acting on an arbitrary state $\ket{\psi(t)}$, it can be expressed as
\begin{equation}\label{Eq:App:Continuum:prop}
	i \hbar \pdv{\ket{\psi(t)}}{t}
		=
		 \Big[
			\dfrac{\hat{p}^2}{2m}
			+ V_{\text{at}}(\hat{r})
			+ \mathsf{e}\hat{r}
			\big(
				E_{\text{cl}}(t)
				+ \hat{E}_L(t)
			\big)
		\Big]\ket{\psi(t)}.
\end{equation}

Among direct ATI and high-order ATI processes, we are particularly interested in the former---those in which the electron is ionized and does not subsequently rescatter with the parent ion.~Following ionization, the electron is accelerated by the strong-laser field, acquiring high velocities and moving far from the nucleus, where the influence of the atomic potential $V_{\text{at}}(\hat{r})$ becomes negligible.~This observation has motivated many early theoretical treatments in strong-field physics to model the post-ionization electron dynamics as those of a free-particle interacting with the field~\cite{keldysh_ionization_1965,faisal_multiple_1973,reiss_effect_1980,lewenstein_theory_1994}.~However, such an approximation is insufficient for describing high-order ATI events, where the electron revisits the atomic core and undergoes rescattering.~This rescattering enables the electron to gain significantly more kinetic energy, giving rise to a secondary plateau at high energies in the photoelectron spectra.~To address these limitations, Ref.~\cite{lewenstein_rings_1995} introduced an extension of the original SFA framework~\cite{lewenstein_theory_1994}, referred to as the \emph{generalized SFA}, in which rescattering events are included perturbatively.~This methodology has since been extended to $S$-matrix approaches~\cite{lohr_above-threshold_1997}, as well as to Coulomb-Distorted SFA models, where the interaction with the atomic potential is non-perturbatively included via Feynmann path-integral techniques~\cite{lai_influence_2015,maxwell_controlling_2016,lai_near-threshold_2017,maxwell_coulomb-corrected_2017,maxwell_analytic_2018,maxwell_coulomb-free_2018}.

Here, in the spirit of Refs.~\cite{lewenstein_rings_1995,lohr_above-threshold_1997}, we consider a perturbative expansion of Eq.~\eqref{Eq:App:Cont:Prop:state} (or, equivalently, Eq.~\eqref{Eq:App:UL:diff:eq}) around the atomic potential. Specifically, we retain only the zeroth-order term, which corresponds to direct ATI events. This in a well justified approximation when working with photoelecton energies around $2 U_p$, with $U_p = E_0^2/(4\omega_L^2)$ the pondemorotive energy, and within the near infrared regime ($\lambda_L \approx 800$ nm). However, it becomes less accurate in the mid-infrared regime ($\lambda_L \approx 2000$ nm), where the electron is more slowly driven away from the parent ion. In such cases, the atomic potential plays a more significant role, giving rise to prominent low energy structures in the photoelectron spectra~\cite{quan_classical_2009,blaga_strong-field_2009} which are not captured within standard SFA. However, for the laser parameters considered here, this zeroth-order treatment remains appropriate, yielding
\begin{equation}\label{Eq:App:Cont:Prop:II}
	i \hbar \pdv{\ket{\psi(t)}}{t}
		=
		\Big[
			\dfrac{\hat{p}^2}{2m}
			+ \mathsf{e}\hat{r}
			\big(
				E_{\text{cl}}(t)
				+ \hat{E}_L(t)
			\big)
		\Big]\ket{\psi(t)},
\end{equation}
which corresponds to the zeroth-order term in a perturbative expansion around $V_{\text{at}}(t)$ in Eq.~\eqref{Eq:App:Cont:Prop:state}.

For reasons that will become clearer in the following analysis, it is convenient to express Eq.~\eqref{Eq:App:Cont:Prop:II} in a velocity-like gauge. This can be achieved via the unitary transformation $\ket{\psi(t)} = \hat{U}_{\text{vg}}(t)\relaxket{\bar{\psi}(t)} = e^{i\mathsf{e}\hat{r}[A_{\text{cl}}(t) + \hat{A}_L(\xi,t)]/\hbar}\relaxket{\bar{\psi}(t)}$, where $A_{\text{cl}}(t)$ is the classical vector potential, related to the electric field through $E_{\text{cl}}(t) = -\pdv*{A_{\text{cl}}(t)}{t}$, and $\hat{A}_L(\xi,t)$ is the vector potential operator, satisfying $\hat{E}_L(\xi,t) = -\partial{\hat{A}_L(\xi,t)}/{\partial t}$. In our case, the vector potential operator takes the explicit form
\begin{equation}
	\begin{aligned}
	\hat{A}_L(\xi,t)
		&=  i \Bigg[
					\hat{a}_L \int \dd t \ f(\xi,t) 
					- \hat{a}_L^\dagger \int \dd t \ f^*(\xi,t)
				\Bigg]
		\\&= i g(\omega_L)
			\Bigg[
				i\bigg(
					\dfrac{\cosh(r)}{\omega} e^{-i\omega t}
					- \dfrac{\sinh(r)}{\omega} e^{i(\omega t - \theta)}
				\bigg)\hat{a}_L
				+ i\bigg(
						\dfrac{\cosh(r)}{\omega} e^{i\omega t}
						- \dfrac{\sinh(r)}{\omega} e^{-i(\omega t - \theta)}
				\bigg)\hat{a}^\dagger_L
			\Bigg]
		\\& \equiv i\big[
								F(\xi,t) \hat{a}_L^\dagger
								- F^*(\xi,t)\hat{a}_L
							\big].
	\end{aligned}
\end{equation}
Using this transformation, Eq.~\eqref{Eq:App:Cont:Prop:II} simplifies to
\begin{equation}\label{Eq:App:Cont:Prop:III}
	i\hbar \pdv{\relaxket{\bar{\psi}(t)}}{t}
		=\dfrac{1}{2m}
			\Big[
				\hat{p}
				+ \mathsf{e}A_{\text{cl}}(t)
				+ \mathsf{e} \hat{A}_L(\xi,t)
			\Big]^2\relaxket{\bar{\psi}(t)},
\end{equation}
which we refer to as $\relaxket{\bar{\psi}(t)} = \hat{U}_V(t,t_0)\relaxket{\bar{\psi}(t_0)}$. Having in mind that $\relaxket{\psi(t)} = \hat{U}_{\text{vg}}(t_0)\relaxket{\bar{\psi}(t_0)}$, we can then write for the original $\ket{\psi(t)}$
\begin{equation}\label{Eq:App:Cont:Prop:state}
	\ket{\psi(t)}
		= \underbrace{e^{i\mathsf{e}\hat{r}[A_{\text{cl}}(t) + \hat{A}_L(\xi,t)]/\hbar}}_{3}
				\underbrace{\hat{U}_V(t,t_0)}_{2}
			\underbrace{e^{-i\mathsf{e}\hat{r}[A_{\text{cl}}(t_0) + \hat{A}_L(\xi,t_0)]/\hbar}}_1
			\ket{\psi(t_0)}.
\end{equation}

The unitary operator given above cannot, in general, be decomposed as $\hat{U}_{\mathsf{e}}(t,t_0) \otimes \hat{U}_{\text{field}}(t,t_0)$, indicating that during the electron's excursion in the continuum, entanglement between light and matter degrees of freedom emerges. The degree of this entanglement is expected to depend strongly on the amount of squeezing. In the limit where $g(\omega_L) \to 0$, the operator reduces to $\hat{U}_L(t,t_0) \to  \hat{U}_{\mathsf{e}}(t,t_0) \otimes \mathbbm{1}$, and the evolution becomes separable. However, when squeezing becomes sufficiently strong---comparable in magnitude to the classical field strength, $E_0 \propto \abs{\alpha}g(\omega_L)$---the operator $\hat{U}_L(t,t_0)$ generates three distinct contributions, denoted as $1$, $2$ and $3$ in Eq.~\eqref{Eq:App:Cont:Prop:state}, each modifying the joint light-matter state immediately following ionization in a different way. About these:
\begin{itemize}
	\item Term 1, when acting on a momentum eigenstate, describes the momentum shift imparted by the electron upon ionization at time $t_1$. In the absence of quantum optical effects---when the light-matter coupling vanishes---its action is simple: it shifts the momentum state as $\ket{v}\to\relaxket{v-\mathsf{e}A_{\text{cl}}(t_1)}$. However, for non-negligible coupling strengths, its action is no longer deterministic and leads to the generation of light-matter entanglement already at the ionization stage. This becomes clear when acting on an initially separable light-matter state
	\begin{equation}\label{Eq:App:Ent:@:ion}
		e^{-i\mathsf{e}\hat{r}(A_{\text{cl}}(t) + \hat{A}(\xi,t))/\hbar}
			\ket{v}\otimes \ket{\Phi}
				= \dfrac{1}{\sqrt{2\pi\hbar}}
				\int \dd x\
					e^{ix[v-\mathsf{e}A_{\text{cl}}(t)]/\hbar}
					\hat{D}\big(\tfrac{e}{\hbar}x F(\xi, t)\big)
						\ket{x}\otimes \ket{\Phi},
	\end{equation}
	which, in general, results in an entangled state, since the displacement imparted to the field depends explicitly on the electron's position.
	
	\item Term 2 governs the joint evolution of the electron and the light field during the electron's propagation in the continuum. As we discuss later, these dynamics result in a momentum-dependent displacement of the field degrees of freedom---stemming from the $[\hat{p}+A_{\text{cl}}(t)]\hat{A}_L(\xi,t)$ contribution---as well as squeezing of the fundamental mode---arising from the $\hat{A}_L^2(\xi,t)$ contribution.
	
	\item Term 3 is structurally similar to term 1, but applies at the later time $t$, associated here with the electron's measurement time.~While $\hat{A}_{\text{cl}}(t)$ and $\hat{A}_L(\xi,t)$ are both periodic with period $T=2\pi/\omega_L$, in general $t \neq t_0 + nT$ for $n\in \mathbbm{N}$.~Therefore, even in the absence of term 2, term 3 does not necessarily cancel the effects induced by term 1.
\end{itemize}

Based on this analysis, we can express the direct ATI (dATI) contribution of Eq.~\eqref{Eq:App:ATI} as
\begin{equation}
	\relaxket{\tilde{\Psi}_{\text{d}}(t)}
		= - \dfrac{i}{\hbar}
				\int^t_{t_0} \dd t_1
					\int \dd k \
						\hat{U}_{\text{vg}}(t)
						\hat{U}_V(t,t_1)\hat{U}^\dagger_{\text{vg}}(t_1)
							\bra{k}
								\big[
									\hat{V}_L(t_1)
									+ \hat{V}_{\text{uv}}(t_1)
								\big]
								\ket{\text{g}}
								e^{iI_p(t_1-t_0)/\hbar}
								\ket{k} \otimes \ket{\bar{0}},
\end{equation}
which becomes more tractable by inserting the identity in the position and momentum representations before the operators $\hat{U}_{\text{vg}}(t_1)$ and $\hat{U}_V(t,t_1)$, respectively. This yields
\begin{equation}\label{Eq:App:state:dATI:I}
	\begin{aligned}
	\relaxket{\tilde{\Psi}_{\text{d}}(t)}
		&= - \dfrac{i\mathsf{e}}{\sqrt{2\pi\hbar^3}}
			\int^t_{t_0} \dd t_1\int \dd v \int \dd x\
				\hat{U}_{\text{vg}}(t)
				\hat{U}_V(t,t_1)
				\hat{D}_L\big(xF(\xi,t_1)\big)
				\big[
					E_{\text{cl}}(t_1)
					+\hat{E}_L(\xi,t_1)
					+ \hat{E}_{\text{uv}}(t)
				\big]
				\\&\hspace{5cm}\times
				e^{i[I_p(t_1-t_0) - x(v+\mathsf{e}A_{\text{cl}}(t_1))]/\hbar}
				h(x)
				\ket{v} \otimes \ket{\bar{0}},
	\end{aligned}
\end{equation}
where we have defined $h(x) \equiv \int \dd k\braket{x}{k}\!\mel{k}{\hat{r}}{\text{g}}$, and made use of Eq.~\eqref{Eq:App:Ent:@:ion}. In the reminder of this section, we elaborate on the structure of $h(x)$ as well as the action of $\hat{U}_V(t,t_1)$ on initial product states of the form $\ket{v}\otimes \ket{\Phi_0}$, where $\ket{\Phi_0}$ denotes an arbitrary quantum optical state. However, it is worth remarking that, when setting $g(\omega_L)\to 0$, the expressions we recover match those found in semiclassical SFA-based analyses of dATI~\cite{lewenstein_rings_1995,lohr_above-threshold_1997,amini_symphony_2019}.

\subsubsection{On the form of $h(x)$}
Now, our aim is to derive an analytical expression for $h(x)$. To simplify the analysis---and motivated by the fact that we are interested in transitions to high-energy continuum states---we assume that the electronic continuum wavefunctions can be approximated by plane waves. Adopting a Gaussian model for the atomic potential, we follow the approximation from Ref.~\cite{lewenstein_theory_1994}
\begin{equation}
	\mel{k}{\hat{r}}{\text{g}}
		\approx
			-i \bigg(
					\dfrac{1}{\pi \alpha}
				\bigg)^{3/4}
				\dfrac{k}{\alpha}
				\exp[-\dfrac{k^2}{2\alpha}],
\end{equation}
where, when working in atomic units, $\alpha = 0.8 I_p$~\cite{nayak_saddle_2019}.~Given that $\braket{x}{k} \approx (2\pi\hbar)^{-1/2}e^{ix k/\hbar}$, we can perform the integral with respect to $k$
\begin{equation}
	h(x)
	= \sqrt{\dfrac{\alpha}{\pi\hbar}}
	\bigg(
	\dfrac{1}{\pi\alpha}
	\bigg)^{1/4}
	\dfrac{x}{\alpha^2}
	\exp[-\dfrac{\alpha x^2}{2\hbar^2}].
\end{equation}

\subsubsection{Continuum states evolution}
Next, we proceed to evaluate the action of the operator $\hat{U}_V(t)$ on states of the form $\ket{v}\otimes \ket{\Phi(t)}$, where $\ket{\Phi(t)}$ denotes an arbitrary pure quantum optical state $\ket{v}$. It is important to note that $\hat{U}_V(t)$ is diagonal in the momentum basis $\ket{v}$. Therefore, by projecting Eq.~\eqref{Eq:App:Cont:Prop:III} onto a momentum eigenstate $\ket{v}$, we obtain
\begin{equation}\label{Eq:App:QO:cont}
	\begin{aligned}
	i \hbar \pdv{\ket{\Phi(v,t)}}{t}
		&= \dfrac{1}{2m}
				\big[
					v
					+ \mathsf{e} A_{\text{cl}}(t)
					+ \mathsf{e}\hat{A}_L(\xi,t)
				\big]^2 \ket{\Phi(v,t)}
		\\&=
			\bigg\{
				\underbrace{\dfrac{1}{2m}
					\big[
						v + \mathsf{e}A_{\text{cl}}(t)
					\big]^2}_{1}
				+ \underbrace{\dfrac{\mathsf{e}}{m} 
					\big[
						v + \mathsf{e}A_{\text{cl}}(t)
					\big] \hat{A}_L(\xi,t)}_{2}
				+ \underbrace{\dfrac{\mathsf{e}^2}{2m}\hat{A}_L^2(\xi,t)}_{3}
			\bigg\} \ket{\Phi(v,t)}
	\end{aligned}
\end{equation}
In the evolution of the quantum optical contribution, we identify three distinct terms:
\begin{itemize}
	\item Term 1 corresponds to the classical contribution of the field to the electron's kinetic energy. It induces a time-dependent phase on the state.
	\item Term 2 represents a linear coupling between the electron's motion and the driving field's creation and annihilation operators. This term leads to a time-dependent displacement of the quantum optical degrees of freedom driven by the electronic dynamics~\cite{rivera-dean_light-matter_2022,stammer_quantum_2023,rivera-dean_role_2024}. In the regime of strong squeezing, where $\cosh(r) \approx \sinh(r) \approx e^{r}/2$, this contribution scales as $g(\omega_L)e^{r}/\omega_L$. In our case, with $g(\omega_L)\sim 10^{-8}$ and $\omega_L\sim 10^{-2}$, the scaling becomes approximately $10^{-6} e^r$.
	\item Term 3, commonly referred to as the diamagnetic term, introduces second-order contributions of the creation and annihilation operators, typically associated to squeezing features. This term scales as $g(\omega_L)^2 e^{2r}/\omega_L^2$, which for our parameters amounts to $10^{-14}e^{2r}$.
\end{itemize}

In our analysis, we focus on regimes where the squeezing introduced in the driving field results in intensities comparable to, or lower than, those of a classical electric field.~That is, we consider $\alpha g(\omega_L) \sim 10^{-2} \geq e^{r}g(\omega_L)$, which implies $10^{6}\geq e^{r}$. Under the most extreme conditions where this inequality becomes an equality, we find that the contribution from Term 2 (scaling as $\sim\!\! 1$) significantly outweighs that of Term 3 (scaling as $\sim\!\! 10^{-2}$). This justifies the approximation that squeezing effects, represented by Term 3, are negligible within the parameter regimes of interest.

As a result, we simplify Eq.~\eqref{Eq:App:QO:cont} by neglecting the contribution of Term 3, yielding
\begin{equation}
	i\hbar \pdv{\ket{\Phi(v,t)}}{t}
		\approx
			\bigg\{
				\dfrac{1}{2m}
					\big[
						v + \mathsf{e}A_{\text{cl}}(t)
					\big]^2
					+ \dfrac{\mathsf{e}}{2m}
						\big[
							v + \mathsf{e}A_{\text{cl}}(t)
						\big]\hat{A}_L(\xi,t)
			\bigg\}\ket{\Phi(v,t)}.
\end{equation}

The solution to this simplified evolution equation, up to a global phase that scales with $[g(\omega_L) e^{r}]^2$, can be expressed as~\cite{rivera-dean_strong_2022,stammer_quantum_2023,rivera-dean_light-matter_2022,rivera-dean_role_2024}
\begin{equation}
	\ket{\Phi(v,t)}
		= e^{-iS_{\text{sc}}(v,t,t_0)/\hbar}
			\hat{D}
				\big(
					\delta(v,t,t_0)
				\big)\ket{\Phi(v,t_0)},
\end{equation}
where we define $S_{\text{sc}}(v,t,t_0) = \frac{1}{2m}\int^{t}_{t_0} \dd \tau [v + \mathsf{e}A_{\text{cl}}(\tau)]^2$ and
\begin{equation}
	\delta(v,t,t_0)
		= \dfrac{\mathsf{e}}{2m\hbar}
			\int^{t}_{t_0} \dd \tau
				\big[
					v + A_{\text{cl}}(\tau)
				\big] F(\xi,\tau).
\end{equation}

\subsubsection{The final state at time $t$}
In the following, we conveniently set the measurement time $t = 2\pi n_{\text{cyc}}/\omega$, where $n_{\text{cyc}} \in \mathbbm{W}$ denotes the number of field cycles.~This choice is motivated by the fact that, at such times, both the semiclassical values of canonical and kinetic momentum coincide. With this, and incorporating the analysis developed in the previous two subsections together with the fact that the effect of $\hat{U}_{\text{vg}}(t_2)$ is alike that of $\hat{U}^\dagger_{\text{vg}}(t_1)$, we can express Eq.~\eqref{Eq:App:state:dATI:I} as
\begin{equation}\label{Eq:App:state:dATI:II}
	\begin{aligned}
		\relaxket{\tilde{\Psi}_{\text{d}}(t)}
			&= - \dfrac{i\mathsf{e}}{\sqrt{2\pi\hbar^3}}
				\int^t_{t_0} \dd t_1 \int \dd x_2\int \dd v \int \dd x_1\					
					\hat{D}_L\big(\delta(v,t,t_1)-\tfrac{e}{\hbar}x_2 F(\xi,t)+\tfrac{e}{\hbar}xF(\xi,t_1)\big)
					\big[
						E_{\text{cl}}(t_1)
						+\hat{E}_L(\xi,t_1)
						+ \hat{E}_{\text{uv}}(t)
					\big]
					\\&\hspace{5cm}\times
						h(x_1)
						e^{-i[S_{\text{sc}}(v,t,t_1) - I_p(t_1-t_0) 
							- x_2v
							+ x_1(v+\mathsf{e}A_{\text{cl}}(t_1))]/\hbar}
							\ket{x_2} \otimes \ket{\bar{0}},
	\end{aligned}
\end{equation}
where the all displacement operators are combined directly, without an additional phase factor, since they share the same phase. 

\subsection{Evaluating the saddle-point equations}\label{Sec:App:SPA}


In the main text, we focus on the analysis of both phase- and amplitude-squeezed states, which in our scheme correspond to setting $\theta = 0$ and $\theta = \pi$, respectively. For phase squeezing, we find that
\begin{equation}
	F(\xi,t)
	= i g(\omega_L)\bigg(
	\dfrac{\cosh(r)}{\omega_L}e^{i\omega_L t}
	- \dfrac{\sinh(r)}{\omega_L}e^{-i\omega_L t}
	\bigg)
	\simeq 
	- \dfrac{e^{r}g(\omega_L)}{\omega_L}	\sin(\omega_L t),
\end{equation}
which when evaluated at $t= 2\pi n_{\text{cyc}}/\omega_L$, yields $F(\xi,t) = 0$. Consequently, the quantum optical state obtained after projection onto a final electron momentum state $\ket{v_{\text{f}}}$ takes the form
\begin{equation}\label{Eq:App:Phi:Phase}
	\begin{aligned}
		\relaxket{\tilde{\Phi}_{\text{d}}(t)}
		&= \langle v_{\text{f}} \vert \tilde{\Psi}_{\text{d}}(t)\rangle
			\\&
		=  - \dfrac{i\mathsf{e}}{\sqrt{2\pi\hbar^3}}
		\int^t_{t_0} \dd t_1 \int \dd x_1\					
		\hat{D}_L\big(\delta(v_{\text{f}},t,t_1)+\tfrac{e}{\hbar}x_1F(\xi,t_1)\big)
		\big[
		E_{\text{cl}}(t_1)
		+\hat{E}_L(\xi,t_1)
		+ \hat{E}_{\text{uv}}(t)
		\big]h(x_1)
		\\&\hspace{4cm}\times
		e^{-i[S_{\text{sc}}(v_{\text{f}},t,t_1) - I_p(t_1-t_0) + x_1(v_{\text{f}}+\mathsf{e}A_{\text{cl}}(t_1))]/\hbar}
		\ket{\bar{0}}.
	\end{aligned}
\end{equation}

In contrast, for the case of amplitude squeezing, we find that the $F(\xi,t)$ function reads
\begin{equation}
	F(\xi,t)
		= i g(\omega_L)
			\bigg(
				\dfrac{\cosh(r)}{\omega_L}e^{i\omega_L t}
				+ \dfrac{\sinh(r)}{\omega_L}e^{-i\omega_L t}
				\bigg)
			\simeq 
			i\dfrac{e^{r}g(\omega_L)}{\omega_L}	\cos(\omega_L t),
\end{equation}
which, unlike phase squeezing case, reaches its a critical point at $t= 2\pi n_{\text{cyc}}/\omega_L$. As a result, the quantum optical state obtained upon projection of the electronic part onto $\ket{v_{\text{f}}}$ involves a greater number of integrals, and is more explicitly given by
\begin{equation}\label{Eq:App:Phi:Amp}
	\begin{aligned}
		\relaxket{\tilde{\Phi}_{\text{d}}(t)}
			&= - \dfrac{i\mathsf{e}}{\sqrt{2\pi\hbar^3}}
			\int^t_{t_0} \dd t_1 \int \dd x_2\int \dd v \int \dd x_1\					
			\hat{D}_L\big(\delta(v,t,t_1)-\tfrac{\mathsf{e}}{\hbar}x_2 F(\xi,t)+\tfrac{\mathsf{e}}{\hbar}x_1F(\xi,t_1)\big)
				\big[
					E_{\text{cl}}(t_1)
					+\hat{E}_L(\xi,t_1)
					+ \hat{E}_{\text{uv}}(t)
				\big]
			\\&\hspace{5cm}\times
			h(x_1)
			e^{-i[S_{\text{sc}}(v,t,t_1) - I_p(t_1-t_0) 
				- x_2(v-v_{\text{f}})
				+ x_1(v+\mathsf{e}A_{\text{cl}}(t_1))]/\hbar}
			\ket{\bar{0}}.
	\end{aligned}
\end{equation}

To evaluate the properties of the state in Eqs.~\eqref{Eq:App:Phi:Phase} and \eqref{Eq:App:Phi:Amp}, we employ the saddle-point approximation, following an approach analogous to that presented in Ref.~\cite{rivera-dean_role_2024}. This approximation is particularly useful for simplifying integrals involving highly oscillatory functions, by approximating the full integral as a sum over dominant contributions at specific points---namely, the saddle-points of the integrand's rapidly varying phase. In our case, however, the integrand includes operators whose action on the initial state explicitly depends on the integration variables. More generally, the quantum state under consideration can be written as
\begin{equation}\label{Eq:App:QO:state:simp}
	\begin{aligned}
	\relaxket{\Tilde{\Phi}_{\text{d}}(t)}
		&= \int \dd \boldsymbol{\theta}
				\hat{D}_L\big(\alpha(\boldsymbol{\theta})\big)
					\big[
						E_{\text{cl}}(t_1)
						+\hat{E}_L(\xi,t_1)
						+ \hat{E}_{\text{uv}}(t)
					\big]h(x_1)
					e^{-i \mathcal{S}(\boldsymbol{\theta})}
					\ket{\bar{0}}
		\\&
		= \int \dd \boldsymbol{\theta}
			\big[
				E_{\text{cl}}(t_1)
				- E_{\alpha(\boldsymbol{\theta})}(\xi,t)
				+\hat{E}_L(\xi,t_1)
				+ \hat{E}_{\text{uv}}(t)
			\big]h(x_1)
		e^{-i \mathcal{S}(\boldsymbol{\theta})}
			\hat{D}_L\big(\alpha(\boldsymbol{\theta})\big)
			\ket{\bar{0}}
		\end{aligned},
\end{equation}
where, in going from the first to the second equality, we have rearranged the displacement operator to act directly on the vacuum state $\ket{\bar{0}}$. To proceed analytically, we expand $\hat{D}_L(\alpha(\boldsymbol{\theta}))\ket{\bar{0}}$ in the Fock basis, where the dependence on the integration variables is transferred to the expansion coefficients, i.e., the probability amplitudes, and the quantum states themselves become independent of the integration variables. This yields
\begin{equation}
	\relaxket{\Tilde{\Phi}_{\text{d}}(t)}
		= \sum_{n=0}^{\infty}
			\int \dd \boldsymbol{\theta}
			\big[
				E_{\text{cl}}(t_1)
				- E_{\alpha(\boldsymbol{\theta})}(\xi,t)
				+\hat{E}_L(\xi,t_1)
				+ \hat{E}_{\text{uv}}(t)
			\big]h(x_1)
			e^{-i \mathcal{S}(\boldsymbol{\theta})- \abs{\alpha(\boldsymbol{\theta})}^2/2}
			\dfrac{\alpha(\boldsymbol{\theta})^n}{\sqrt{n!}}
		\ket{n} \bigotimes_{q\neq 1} \relaxket{0_q}.
\end{equation}

In this way, all the dependence on the integration variables is transferred to the probability amplitudes, which are amenable to the saddle-point approximation. However, we emphasize that that the saddle-point equations---and the number of saddles---differ between the phase- and amplitude-squeezed cases. Specifically, for phase squeezing the function used to identify the saddle points is given by
\begin{equation}
	S_{\text{QO}}^{(\text{ph})}(\boldsymbol{\theta})
		= S_{\text{cl}}(v,t,t_1) - I_p(t_1-t_0)
		- i \hbar \dfrac{\abs{\alpha_{\text{ph}}(\boldsymbol{\theta})}^2}{2}
		- \dfrac{i}{2}
		x_1
		\big[
		i2\big(v_{\text{f}} + \mathsf{e}A_{\text{cl}}(t_1)\big)
		+ x_1 \alpha
		\big],		
\end{equation}
where $\alpha_{\text{ph}}(\boldsymbol{\theta}) \equiv \delta(v_{\text{f}},t,t_1)+\tfrac{e}{\hbar}x_1F(\xi,t_1)$.~The corresponding saddle-point equations are obtained by setting $\nabla_{\boldsymbol{\theta}} S_{\text{QO}}^{(\text{ph})}(\boldsymbol{\theta})\vert_{\boldsymbol{\theta}_s} = \boldsymbol{0}$, where $\boldsymbol{\theta}_s = (t_{\text{ion}},x_{1,s})$ denotes the saddle-point coordinates.~More explicitly, the saddle-point equations are given by
\begin{align}
	&\pdv{S_{\text{QO}}^{(\text{ph})}(\boldsymbol{\theta})}{t_1}\Big|_{\boldsymbol{\theta}_s} = 0
	\Rightarrow
		\dfrac{[v+\mathsf{e}A_{\text{cl}}(t_{\text{ion}})]^2}{2m}
		+ I_p -x_{1,s} E_{\text{cl}}(t_{\text{ion}})
		- \dfrac{i\hbar}{2}
			\pdv{\abs{\alpha_{\text{ph}}(\boldsymbol{\theta})}^2}{t_1}\Big|_{\boldsymbol{\theta}_s} = 0, \label{Eq:App:saddle:time:ph}
	\\&\pdv{S_{\text{QO}}^{(\text{ph})}(\boldsymbol{\theta})}{x_1}\Big|_{\boldsymbol{\theta}_s} = 0
	\Rightarrow
		\big(
			v_{\text{f}}
			+ \mathsf{e} A_{\text{cl}}(t_{\text{ion}})
		\big)
		-i\alpha x_{1,s}
		- \dfrac{i\hbar}{2}
		\pdv{\abs{\alpha_{\text{ph}}(\boldsymbol{\theta})}^2}{x_1}\Big|_{\boldsymbol{\theta}_s} = 0.\label{Eq:App:saddle:x1:ph}
\end{align}

In contrast, for amplitude squeezing, the function used to determine the saddle-points is given by
\begin{equation}
	S_{\text{QO}}^{(\text{ph})}(\boldsymbol{\theta})
		= S_{\text{cl}}(v,t,t_1) - I_p(t_1-t_0)
			- i \hbar \dfrac{\abs{\alpha_{\text{amp}}(\boldsymbol{\theta})}^2}{2}
			- \dfrac{i}{2}
				x_1
				\big[
					i2\big(v + \mathsf{e}A_{\text{cl}}(t_1)\big)
					+ x_1 \alpha
				\big]
			-x_2(v-v_{\text{f}}),		
\end{equation}
where $\alpha_{\text{amp}}(\boldsymbol{\theta}) \equiv \delta(v_{\text{f}},t,t_1)-\tfrac{e}{\hbar}x_2F(\xi,t)+\tfrac{e}{\hbar}x_1F(\xi,t_1)$. The corresponding saddle-point equations, with solutions given in this case by $\boldsymbol{\theta}_s = (t_{\text{ion}},x_{2,s},x_{1,s},v_{s})$, are as follows
\begin{align}
	&\pdv{S_{\text{QO}}^{(\text{amp})}(\boldsymbol{\theta})}{t_1}\Big|_{\boldsymbol{\theta}_s} = 0
	\Rightarrow
		\dfrac{[v+\mathsf{e}A_{\text{cl}}(t_{\text{ion}})]^2}{2m}
			+ I_p -x_{1,s} E_{\text{cl}}(t_{\text{ion}})
			- \dfrac{i\hbar}{2}
			\pdv{\abs{\alpha_{\text{amp}}(\boldsymbol{\theta})}^2}{t_1}\Big|_{\boldsymbol{\theta}_s} = 0,\label{Eq:App:saddle:time:amp}
	\\&\pdv{S_{\text{QO}}^{(\text{amp})}(\boldsymbol{\theta})}{x_1}\Big|_{\boldsymbol{\theta}_s} = 0
	\Rightarrow
		\big(
			v_{\text{f}}
			+ \mathsf{e} A_{\text{cl}}(t_{\text{ion}})
		\big)
		-i\alpha x_{1,s}
		- \dfrac{i\hbar}{2}
	\pdv{\abs{\alpha_{\text{amp}}(\boldsymbol{\theta})}^2}{x_1}\Big|_{\boldsymbol{\theta}_s} = 0,
	\\&\pdv{S_{\text{QO}}^{(\text{amp})}(\boldsymbol{\theta})}{x_2}\Big|_{\boldsymbol{\theta}_s} = 0
		\Rightarrow
			(v_s-v_{\text{f}}) + \dfrac{i\hbar}{2}
			\pdv{\abs{\alpha_{\text{amp}}(\boldsymbol{\theta})}^2}{x_2}\Big|_{\boldsymbol{\theta}_s} = 0,
	\\&\pdv{S_{\text{QO}}^{(\text{amp})}(\boldsymbol{\theta})}{v}\Big|_{\boldsymbol{\theta}_s} = 0
		\Rightarrow
		\dfrac{1}{m}
		\int^{t}_{t_{\text{ion}}}
			\!\!\dd \tau 
				\big[
					v_s + \mathsf{e}A_{\text{cl}}(\tau)
				\big]
				- (x_{2,s}-x_{1,s})
				- \dfrac{i\hbar}{2}
					\pdv{\abs{\alpha_{\text{amp}}(\boldsymbol{\theta})}^2}{x_2}\Big|_{\boldsymbol{\theta}_s} = 0.\label{Eq:App:saddle:v:amp}
\end{align}

To obtain solutions of both the phase squeezing [Eqs.~\eqref{Eq:App:saddle:time:ph}-\eqref{Eq:App:saddle:x1:ph}] and the amplitude squeezing equations [Eqs.~\eqref{Eq:App:saddle:time:amp}-\eqref{Eq:App:saddle:v:amp}] for different values of $v_{\text{f}}$ and $n_{\text{cyc}}$, we employ numerical root-finding techniques.~Specifically, we use the \texttt{FindComplexRoots} solver implemented in the \texttt{RBSFA} Mathematica package~\cite{RBSFA}.

After applying the saddle-point approximation, we can generally rewrite Eq.~\eqref{Eq:App:QO:state:simp} as
\begin{equation}
	\relaxket{\Tilde{\Phi}_{\text{d}}(t)}
	= \sum_{\boldsymbol{\theta}_s}G(\boldsymbol{\theta}_s)
	\hat{D}_L\big(\alpha(\boldsymbol{\theta}_s)\big)
		\big[
			E_{\text{cl}}(t_1)
			+\hat{E}_L(\xi,t_1)
			+ \hat{E}_{\text{uv}}(t)
		\big]
		\ket{\bar{0}},
\end{equation}
with $G(\boldsymbol{\theta}_s)$ a comples-valued prefactor, that arises from evaluating the integrand at the saddle-points. It also accounts for additional weights, depending on the diagonal elements Hessian of the action evaluated at the saddle-points, arising from the application of the saddle-point approximation~\cite{lewenstein_theory_1994,olga_simpleman,nayak_saddle_2019}.

\section{QUANTUM OPTICAL ANALYSIS}

\subsection{Invariance of the negativity volume}\label{Sec:App:QO}
In our analysis, we adopt the Wigner negativity volume as a quantitative measure of non-classicality, specifically employing the definition introduced in Ref.~\cite{kenfack_negativity_2004}. A key advantage of this measure lies in its invariance under Gaussian operations~\cite{walschaers_non-gaussian_2021}, which comparably simplifies numerical calculations. In particular, this property allows us to avoid explicitly implementing the strong squeezing operator associated with the driving field when evaluating the negativity volume (see Sec.~\ref{Sec:App:Numerics}). The purpose of this subsection is to explicitly demonstrate this invariance.

Following Ref.~\cite{royer_wigner_1977}, the Wigner function associated with a quantum state $\hat{\rho}$ can be written as
\begin{equation}
	W(\beta)
		= \tr[
				\hat{D}(-\beta)
					\hat{\Pi}
				\hat{D}(\beta)\hat{\rho}
				]
		= \tr[
			\hat{\Pi}
			\hat{D}(\beta)\hat{\rho}\hat{D}(-\beta)
			],
\end{equation}
where $\hat{\Pi}$ denotes the parity operator. In transitioning from the first to the second equality, we make use of the cyclic property of the trace. In our case, by undoing the squeezing transformation that leads to Eq.~\eqref{Eq:App:Squeez:E}, the state can be written as $\hat{\rho} = \hat{S}(\xi)\hat{\sigma}\hat{S}^\dagger(\xi)$. Substituting this into the expression for the Wigner function yields
\begin{equation}
	W(\beta)
		= \tr[\hat{\Pi}\hat{D}(\beta)\hat{S}(\xi)
					\hat{\sigma}
				\hat{S}^\dagger(\xi)\hat{D}^\dagger(\beta)].
\end{equation}

Taking into account that the displacement and squeezing operators satisfy the identity $\hat{D}(\beta)\hat{S}(\xi) = \hat{S}(\xi)\hat{D}(\gamma)$, with $\gamma = \beta \cosh(\xi) + \beta^* \sinh(\sigma)$ and assuming without loss of generality that $\xi > 0$, we can rewrite the expression above as
\begin{equation}
	W(\beta)
		= \tr[\hat{\Pi}\hat{S}(\xi)\hat{D}(\gamma)\hat{\sigma}\hat{D}^\dagger(\gamma)\hat{S}^\dagger(\xi)].
\end{equation}
Since $\hat{S}(\xi)$ is an even operator---being quadratic in creation and annihilation operators---it commutes with the parity operator $\hat{\Pi}$. Applying the cyclic property of the trace then yields
\begin{equation}
	W(\beta)
	= \tr[
			\hat{S}(\xi)\hat{\Pi}
				\hat{D}(\gamma)\hat{\sigma}\hat{D}^\dagger(\gamma)
			\hat{S}^\dagger(\xi)]
	= \tr[\hat{D}(\gamma)\hat{\sigma}\hat{D}^\dagger(\gamma)]
	= W(\gamma),
\end{equation}
i.e., the Wigner function is simply evaluated at the transformed phase-space point $\gamma$. Summing over all values of $\beta$, we can express the result $N$ as 
\begin{equation}\label{Eq:App:wbeta:wgamma}
	N = \int \dd^2 \beta
				W(\beta)
		= \int \dd^2 \beta W(\gamma)
		= \int \dd^2\gamma W(\gamma)
\end{equation}
where in going from the second to the third equality we used the fact that $\dd^2 \beta = \dd^2 \gamma$. Thus, as expected, the squeezing operation preserves the norm (or total integral) of the Wigner function.

In our case, we are particularly interested in whether $\int \dd \beta \lvert W(\beta)\rvert$ remains invariant under squeezing. This follows straightforwardly from Eq.~\eqref{Eq:App:wbeta:wgamma}, such that we can write
\begin{equation}
	\lvert W(\beta)\rvert
		= \lvert W(\gamma) \rvert.
\end{equation}
Thus, when integrating over $\beta$ while making the change of variables $\beta \to \gamma$ in the right hand side, we arrive at
\begin{equation}
	\int \dd \beta
		\lvert W(\beta)\rvert
	=
	\int \dd \gamma
		\lvert W(\gamma)\rvert,
\end{equation}
where the left-hand side corresponds to the Wigner function of the squeezed state $\hat{S}(\xi)\hat{\sigma}\hat{S}^\dagger(\xi)$, while the right-hand side corresponds to that of the unsqueezed state $\hat{\sigma}$. We conclude that the volume of Wigner negativity is invariant under squeezing.

\subsection{Numerical analysis}\label{Sec:App:Numerics}
The analysis of the quantum optical and quantum information measures was carried out entirely in Python, utilizing the \texttt{QuTiP} package~\cite{johansson_qutip_2012,johansson_qutip_2013}.~Within this framework, the quantum optical states are represented in the Fock basis, with a Hilbert space truncation at 200 elements.~This cutoff was benchmarked against higher values (up to 300) to ensure numerical convergence.~In this context, the use of measures that are invariant under displacement and squeezing operators---such as the negative volume---proved essential.~Without such invariance, the effective Hilbert space dimension required for accurate computation would increase significantly, resulting in substantial additional memory usage and computational time.

\end{document}